\documentclass[twocolumn,superscriptaddress,nofootinbib]{revtex4-1}

\usepackage{comment}
\usepackage{slashed}
\usepackage{latexsym}
\usepackage{graphics}
\usepackage{bm}
\usepackage{color}
\usepackage{amsmath}
\usepackage{amssymb}
\usepackage{graphicx}
\usepackage{subcaption}
\usepackage{slashed}

\begin{document}

\title{Earth-bound Milli-charge Relics}

\author{Maxim Pospelov}
\affiliation{School of Physics and Astronomy, University of Minnesota, Minneapolis, MN 55455, USA}
\affiliation{William I. Fine Theoretical Physics Institute, School of Physics and Astronomy,
University of Minnesota, Minneapolis, MN 55455, USA}

\author{Harikrishnan Ramani}\email{hramani@stanford.edu}
\affiliation{Stanford Institute for Theoretical Physics,
Stanford University, Stanford, CA 94305, USA}

\begin{abstract}

Dark sector particles with small electric charge, or millicharge, (mCPs) may lead to a variety of diverse phenomena in particle physics, astrophysics and cosmology. Assuming their possible existence, we investigate the accumulation and propagation of mCPs in matter, specifically inside the Earth. Even small values of millicharge lead to sizeable scattering cross sections on atoms, resulting in complete thermalization, and as a consequence, considerable build-up of number densities of mCPs, especially for the values of masses of GeV and higher when the evaporation becomes inhibited. Enhancement of mCP densities compared to their galactic abundance, that can be as big as $10^{14}$, leads to the possibility of new experimental probes for this model. The annihilation of pairs of mCPs will result in new signatures for the large volume detectors (such as Super-Kamiokande). Formation of bound states of negatively charged mCPs with nuclei can be observed by direct dark matter detection experiments. A unique probe of mCP can be developed using underground electrostatic accelerators that can directly accelerate mCPs above the experimental thresholds of direct dark matter detection experiments. 

\end{abstract}

\maketitle

\section{Introduction}
Charge quantization is a century old mystery. While explanations for quantization exist, the resultant predictions of magnetic monopoles and/or manifestation of grand unification (GUT) have not been observed despite systematic efforts. This has led to the more open-minded approach to charge quantization, and exploration of the possible existence of non-quantized charges also referred to as milli-charge particles (mCPs). In recent years mCPs have received further theoretical and experimental scrutiny (see {\em e.g.} a selection of papers on theoretical and experimental efforts: \cite{Haas:2014dda,Izaguirre:2015eya,Magill:2018tbb,Kelly:2018brz,
Harnik:2019zee,Acciarri:2019jly,Plestid:2020kdm,Bloch:2020uzh,Foroughi-Abari:2020qar,Harnik:2020ugb,Marocco:2020dqu}). 

On the theoretical side, models with pure mCPs as well as models where smallness of effective electric charge is achieved via photon mixing with a new nearly massless gauge boson have been considered
\cite{Holdom:1985ag}. Since their stability is guaranteed by their $U(1)$ charge, a non-trivial relic abundance surviving from the Big Bang can be expected. Depending on their mass and charge, they could explain all or part of the observed dark matter, called milli-charge dark matter or mCDM with their abundance set by the freeze-out or freeze-in. (Freeze-out refers to the self-depletion through annihilation from the initially fully thermally excited abundance, while the freeze-in is a sub-Hubble-rate-induced population corresponding to smaller couplings.)  Regardless of cosmological abundance of mCPs, there exists a smaller yet irreducible abundance arising from the interaction of cosmic-rays with intervening matter \cite{Plestid:2020kdm,Harnik:2020ugb}. 

Owing to the enhancement of mCP scattering cross sections at low momentum transfer, they have been invoked recently as an explanation of certain low-energy anomalies, such as enhanced absorption of CMB by 21cm absorbers \cite{Barkana:2018qrx,Boddy:2018wzy,Liu:2019knx}, and excess of the keV scale ionization in the Xenon 1T experimental results \cite{Aprile:2020tmw,Farzan:2020dds,Harnik:2020ugb}. 

Regardless of possible anomalous results explained by mCPs, there have been a plethora of
efforts looking for mCPs in collider and beam-dump experiments, that should be viewed in a broader context of exploring the dark sectors \cite{Lanfranchi:2020crw}. mCP relics depending on their speed could also be detected in dark matter direct detection and neutrino experiments. In addition, there are  strong limits on mass vs coupling  parameter space arising from cosmology \cite{Kovetz:2018zan, Creque-Sarbinowski:2019mcm,Slatyer:2018aqg} and galactic astrophysics \cite{Stebbins:2019xjr, Kachelriess:2020ams} as well as from stellar energy losses \cite{Davidson:2000hf, Chang:2018rso}. 

Despite these efforts, there is a tantalizing window of parameter space that current and future experimental efforts cannot access. This window corresponds to $m_Q \approx 10$ MeV and heavier, where BBN bounds do not apply (notice, however that BBN can still limit such models through the excess abundance of dark photons in corresponding models, see {\em e.g.} \cite{Vogel:2013raa}). Defining the mCP charge as $\epsilon e$, with $e$ the electric charge, $ \epsilon \lesssim 0.1$ are not directly limited by collider and beam dump experiments, for $m_Q \approx 1$ GeV and heavier. If these mCPs make up a fraction $f_Q$ of the DM, then for large enough charge, the atmospheric or rock overburden is enough to slow them down to small values of kinetic energies and making them inaccessible to current direct detection (DD) experiments.

In this paper, the main point to be investigated and exploited for possible novel signatures is mCPs slowing down inside the Earth, resulting in a dramatic increase of their number densities at the locations of underground laboratories. This paves the way to novel methods of searching for mCP that we also explore in this paper. 
A direct consequence of the mCP's precipitous slow-down is that this mCP thermalizes with the atmosphere (earth) and for large enough $m_Q$, does not possess a large enough velocity to escape the planet subsequently. 
Barring subsequent evaporation, this builds up through the age of the Earth $t_{\oplus}\sim O(4\times 10^{9})$ years leading to terrestrial densities of mCPs several orders of magnitude larger than the virial density of weakly-interacting DM. If the incoming mCP flux makes up a fraction $f_Q$ of the incoming DM flux, then terrestrial densities as high as $n_Q^{\rm terr} \approx f_Q \frac{10^{14}}{\textrm{cm}^3}$ can be obtained. Depending on the precise value of the mass, this tremendous density increase may be concentrated inside the Earth's core, or be spread out through the whole Earth's volume. Even in the case of heavy masses, the constant vertical downward drift of thermalized mCPs is slow, leading to the ``traffic jam" effect that we have discussed earlier \cite{Pospelov:2019vuf, Lehnert:2019tuw, Rajendran:2020tmw}.

Previous literature has explored the build up of this large density of  DM that has large cross-section with the SM nuclei \cite{Neufeld:2018slx}, as well as its consequences on Earth, stars\,\cite{Busoni:2017mhe}, comets\,\cite{DeLuca:2018mzn} and even exoplanets \cite{Leane:2020wob}. However to our knowledge, the build up of milli-charges specifically and its consequences has missed scrutiny. In this work, we explore various sources of mCP flux on Earth and the subsequent build-up terrestrially. It was shown previously that masses above a GeV sink to the center of the Earth and below a GeV evaporate away leading to a narrow window of mass where this terrestrial accumulation is relevant for experiments near the surface \cite{Neufeld:2018slx}. However, due to the massless-mediator nature of mCP-SM interactions, the mCP slow-down increases this cross-section leading to larger abundances expanding the mCP masses that accumulate appreciably near the Earth's surface. 

The large enhancement in local density compared to the virial density by extremely large factors ({\em e.g.} for some parts of the parameter space by as much as $10^{14}$), opens up novel detection strategies. In this paper, we discuss only a small subset of possible new phenomena and strategies for mCP detection:
\begin{itemize}

\item[-] Terrestrial mCPs with relatively large charge can bind with SM nuclei which can be looked for in exotic isotope searches. If the binding energy is large enough, there is a chance of seeing negative mCP - atomic nucleus ``recombination" inside a detector ({\em cf.} Refs. \cite{Pospelov:2008qx,An:2012bs,Fornal:2020bzz} for analogous ideas).  

\item[-]  For small enough $\epsilon$, where binding is not allowed, annihilations of mCPs with anti-particles into SM can be looked for in the large volume neutrino detectors, and specifically in super-Kamiokande. 

\item[-] Milli-charged particles that have accumulated inside electrostatic accelerators
can be ``accidentally" accelerated (for large enough $Q$) and gain energy. The subsequent
scattering in the low-threshold  direct detection experiments is capable of providing very strong sensitivity 
to the mCPs. This is especially relevant in light of new efforts to install MeV-scale accelerators (for the studies of rare nuclear reactions) in the underground laboratories \cite{Broggini:2017cez}. 

\end{itemize}

The last idea on this list is somewhat reminiscent of the proposal \cite{Berlin:2019uco} that seeks to perturb the flow of mCPs by EM fields, with subsequent detection of this perturbation in the adjacent spatial region. The proposal of Ref. \cite{Berlin:2019uco} is aimed at smaller mass and smaller $\epsilon$ compared to those explored in this paper. 

The rest of this work is organized as follows. 
In Section.~\ref{vdm}, we explore the capture of virial mCDM, subsequent evaporation and the resultant density near the surface. In Section.~\ref{semirel}, sources of high rigidity mCP fluxes are explored and their resultant density near the surface is calculated. This is followed by Section.~\ref{boundstates} which deals with bound state formation between mCPs and SM particles. Section.~\ref{annihilations} deals with the detectable consequences of  mCPs annihilating inside volumes of neutrino detectors. Section.~\ref{accelerator} provides a novel proposal to detect mCPs in an electrostatic accelerator. We present concluding remarks in Section.~\ref{conclusion}. 

\section{Top-Down Accumulation}
\label{vdm}
\begin{figure}[htpb]
\centering
\includegraphics[width=\linewidth]{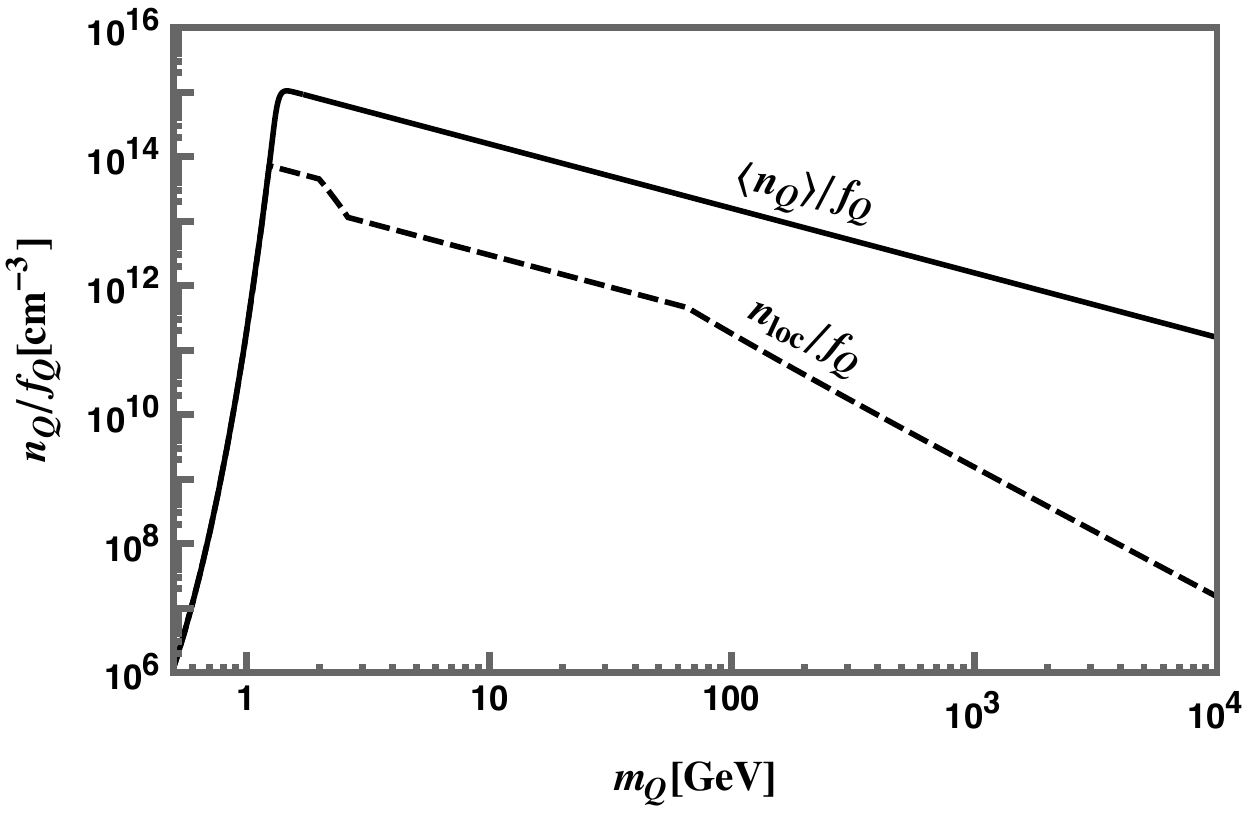} 
\label{fig:dens1}
\caption{Density of millicharge particles $m_Q$ on Earth. The solid line corresponds to the volume averaged equilibrium number density $\langle n_Q \rangle$. Due the Earth's gravity, larger masses sink faster into the earth and only a small fraction of $\langle n_Q \rangle$ is accessible near the surface. The dashed line corresponds to the local number density $n_{\rm loc}$ for large enough millicharge $\epsilon$.}
\label{fig1}
\end{figure}
The goal of this subsection is to consider, in broad strokes, the accumulation of virial mCDM with large enough charge such that thermalization on Earth is rapid. We only consider asymmetric DM in this subsection, such that annihilations can be ignored. The symmetric case where annihilations are relevant for resultant terrestrial densities, is considered in Section.~\ref{annihilations}. 
 
In the literature, two types of mCPs have been considered. The first type is minimal mCPs, truly milli-charged under the SM $U(1)$, without photon-dark-photon mixing. These particles have properties identical to SM charge at all length scales. This property results in highly complex dynamics for the mCP propagation through the galaxy, through the solar system and on Earth. This is because these mCPs interact with the galactic magnetic field, with the solar wind and finally with the electric field between the ionosphere and ground. We leave this problem for future work.

Here, we instead consider particles $Q$ that are charged under a dark U(1) with charge $g_Q$, with the dark photon kinetically mixing with the SM photon with mixing parameter $\kappa$. At energy scales $\omega \gg m_{A'}$, the mass of this dark photon, these particles $Q$ act effectively as mCPs with charge $\epsilon e = \kappa g_Q$, where $e$ is the electron charge. A large enough $m_{A'}$ can be chosen so as to turn off the long range effects described above and yet keep the milli-charge properties which lead to testable consequences that we highlight below. It is likely that such a set-up would lead to tensions with cosmology via the increase of $N_{eff}$ \cite{Vogel:2013raa}, which perhaps could be circumvented at the cost of adding additional ingredients to the evolution of the primordial universe.

\subsection{Capture and evaporation}

For the couplings we are interested in, all of the mCDM gets captured, or in other words, it is in the  multiple scattering regime, $\sigma \times ( n_{\rm atom} \ell)\gg 1$, and will lose its initial kinetic energy down to characteristic thermal energy. The Earth volume averaged number density of dark matter captured is 
\begin{align}
\label{start}
\langle n^{\rm cap}_Q \rangle&= \frac{\pi R_{\oplus}^2 v_{\rm vir}  t_{\oplus}}{4/3 \pi R_{\oplus}^3}f_Q \frac{\rho_{\rm DM}}{m_Q} \nonumber \\
&\approx \frac{3\times 10^{15} }{\textrm{cm}^3} \frac{t_{\oplus}}{10^{10} \textrm{year}} f_Q \frac{\rm GeV}{m_Q}
\end{align}
Here $f_Q$ is the fraction of virialized DM in milli-charges, defined as
\begin{equation}
f_Q = \frac{m_Qn_Q}{\rho_{\rm DM}},
\end{equation}
where $\rho_{\rm DM}$ is the local dark matter density, $\rho_{\rm DM}\simeq 0.3$\,GeV/cm$^3$. 

$v_{\rm vir}$ refers to an average velocity of galactic mCPs. 
Eqn (\ref{start}) neglects gravitational focussing, and in the case of the Earth's capture it is well-justified. 

For smaller masses, dark matter that thermalizes with atmosphere (water, rock etc) 
has a thermal velocity $v_{\rm th} > v_{\rm esc}$. Therefore, there exists a ``last scattering surface" somewhere in the atmosphere, from which the most velocitized mCPs can freely escape, {\em i.e.} evaporate. Adopting earlier results, see {\em e.g.}  \cite{Neufeld:2018slx},  the evaporation rate per one mCP particle  can be estimated as,
\begin{equation}
\Gamma_{\rm loss}\approx \frac{3v_{\rm  th}}{2\pi^\frac{1}{2}R_{\oplus}} \left(1+\frac{v_{\rm es}^2}{v_{\rm  th}^2}\right)\exp\left(-\frac{v_{\rm es}^2}{v_{\rm th}^2}\right)
\end{equation}

The equilibrium density on Earth is given by,
\begin{equation}
\langle n_Q \rangle=\langle n^{\rm cap}_Q \rangle \frac{1-\exp\left(\Gamma_{\rm loss}  t_{\oplus}\right)} {\Gamma_{\rm loss}  t_{\oplus}}
\label{avgdens}
\end{equation}
This is plotted as the solid curve in Fig.~\ref{fig1}. One can see that above 1\,GeV the evaporation is no longer a factor, and the captured number density displays the familiar $m_Q^{-1}$ scaling. It is evident from this plot that the the accumulated density through the lifetime of the Earth can be up to fifteen orders of magnitude larger than the galactic density of the mCPs. 

At the same time, the regime of light mCPs experiencing strong evaporation is more difficult to analyze. In particular, slow-down in the upper atmosphere and subsequent diffusion and evaporation can be altered by many effects including the macroscopic mass transport. Precise analysis of this regime ({\em e.g.} $m_Q\simeq 10-500\,{\rm MeV}$) goes beyond the scope if this work and is excluded from Fig.~\ref{fig1}. 

\subsection{Density near the surface}
While the number density averaged over the Earth volume is given in Eqn.~\ref{avgdens}, the equilibrium density profile as a function of depth depends on the mass of the mCDM. The presence of gravity as well as pressure and temperature gradients results in rearrangement with a density profile that is mass dependent. This profile was evaluated in \cite{Neufeld:2018slx} for the resultant stable population of strongly interacting particles. The main conclusions of \cite{Neufeld:2018slx} were that, for relevant cross-sections, the number density at the surface $n_{\rm jeans}\approx \langle n_Q \rangle$ at $m_Q \le 1$~GeV, while there was diminishing number density of dark matter near the surface for $m_Q \ge 1$ GeV owing to sinking to a greater depth. 

However this sinking is not immediate. Diffusion rates and terminal velocities determine the net sinking of heavier dark matter to lower altitudes. To estimate these rates we need the transfer cross section in terrestrial medium (which we will call ``rock"). The transfer cross-section $\sigma_T$ for thermalized dark matter with atoms is estimated in Appendix A. In the perturbative regime, one can get good estimates with simple models of the charge distribution inside an atom. We find that to a very good approximation for both attractive and repulsive interactions,
\begin{align}
 \sigma_T&\sim \textrm{Min} \left( \frac{2 \pi Z^2 \alpha^2 \epsilon^2}{\mu_{\rm rock,Q}^2 v_{\rm th}^4} , \frac{4 \pi}{\mu_{\rm rock,Q}^2 v_{\rm th}^2} \right)
\label{transfer}
\end{align}
In this formula, $\mu_{\rm rock,Q}^2$ stands for the reduced mass of an atom-mCP system, $Z$ is the atomic number and $v_{\rm th}$ is the typical thermal velocity of a particle with mass $\mu$ . 

While all of the dark matter is captured in the atmosphere or close to the solid Earth's surface, the random walk due to thermal motion will cause DM to diffuse deeper into rock. The time taken to diffuse to depth $h$ is given by,
\begin{equation}
t_{\rm diff} (h)\sim \frac{h^2 n_{\rm rock} \sigma_T}{ v_{\rm th}},
\label{tdiff}
\end{equation}
assuming simple Brownian motion. Here $n_{\rm rock}$ is a typical number density of atoms. 
If $t_{\rm diff} \le t_{\oplus}$, the age of the Earth, this causes the DM to spread out all over the Earth volume and reach the average densities given in Eqn.~\ref{avgdens}. However, in the presence of gravity there is a mechanism for vertical sinking of DM, through the gravitational pull.

Under the gravitational interactions mCPs can acquire terminal velocity that can be estimated as \cite{landau1981course},
\begin{align}
    v_{\rm term} &= \frac{3 m_Q g T}{m_{\rm rock}^2 n_{\rm rock} \langle \sigma_T v_{\rm th}^3 \rangle} 
    \quad \quad && m_Q> m_{\rm rock} \nonumber \\
     &= \frac{m_Q g }{3 n_{\rm rock} T}\left\langle \frac{v_{\rm th}}{\sigma_T} \right\rangle 
      \quad \quad && m_Q < m_{\rm rock}
      \label{term}
\end{align}
While this causes suppression of DM densities near the surface relative to the volume averaged number density in Eqn.~\ref{avgdens}, there is still a traffic jam effect on the way down as pointed out in ~\cite{Pospelov:2019vuf} in the limit $ v_{\rm term} \ll v_{\rm vir}$.
The enhancement in number density occurs because of flux conservation, as smaller velocity at a depth $h$ translate to larger number densities: 
\begin{equation}
\eta_{\rm tj}\equiv \frac{n_{\rm tj}}{n_{\rm vir}}=\frac{v_{\rm vir}}{v_{\rm term}}
\end{equation}
Note that this enhancement should never exceed the volume average in Eqn.~\ref{avgdens}, after all, diffusion will always be more efficient. 
This enhancement factor is the most relevant for the heavy dark matter that has its equilibrium position much closer to the center of the Earth than to its surface. However, for the GeV-scale dark matter, this type of ``traffic jam" enhancement can be still small compared to equilibrium Jeans-type contribution. 
Putting these two enhanced populations together, finally, the number density for $m_Q \ge 1$ GeV at an underground laboratory can be estimated as,
\begin{equation}
n_{\rm loc} = \textrm{Max}\left( n_{\rm jeans}, \textrm{Min}\left( n_{\rm tj} , \langle n_Q \rangle \right)\right)
\label{regimes}
\end{equation}
while for $m_Q < 1$~GeV it is simply given by  $n_{\rm jeans}$ calculated in \cite{Neufeld:2018slx}. The dashed line in Fig.~\ref{fig1} illustrates this total enhanced density as a function of DM mass $m_Q$. This is applicable only for the millicharge $\epsilon$ large enough such that it stops in the corresponding overburden. Thus, this line is applicable only for
\begin{align}
\epsilon&\gtrsim 2\times10^{-4} \sqrt{\frac{m_Q}{\rm GeV}} \quad &&\textrm{surface} \nonumber \\
&\gtrsim 3\times 10^{-6}\sqrt{\frac{m_Q}{\rm GeV}} \quad &&\textrm{1 km mine} 
\end{align}
For such large $\epsilon$, the transfer cross-section is given by the second term in Eqn.~\ref{transfer} which is independent of $\epsilon$ and hence this line is applicable for all $\epsilon$ in this range. One notices three distinct regimes. For $m_Q \lesssim 4$~GeV, the density is dominated by $n_{\rm jeans}$. For larger masses, there is sinking, and the number density is instead given by $n_{\rm tj}$. The kink at around 50 GeV, corresponds to shift from $m_Q \le m_{\rm rock}$ to $m_Q > m_{\rm rock}$ in Eqn.~\ref{term}. 
\section{Bottom-Up Accumulation}
\label{semirel}
\begin{figure*}[htpb]
\centering
    \begin{subfigure}[t]{0.495\textwidth}
        \centering
        \includegraphics[width=\linewidth]{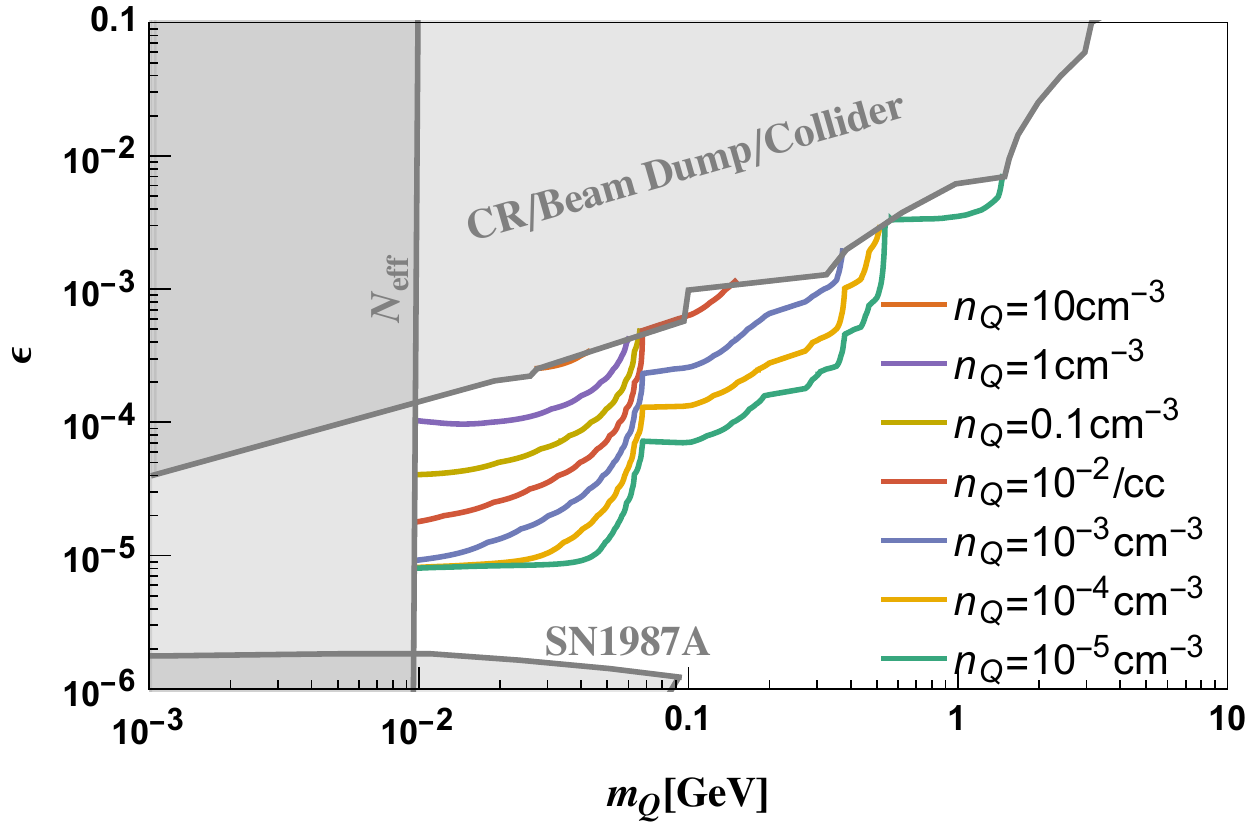} 
        \label{fig:crcap1}
    \end{subfigure}
    \hfill
    \begin{subfigure}[t]{0.495\textwidth}
        \centering
        \includegraphics[width=\linewidth]{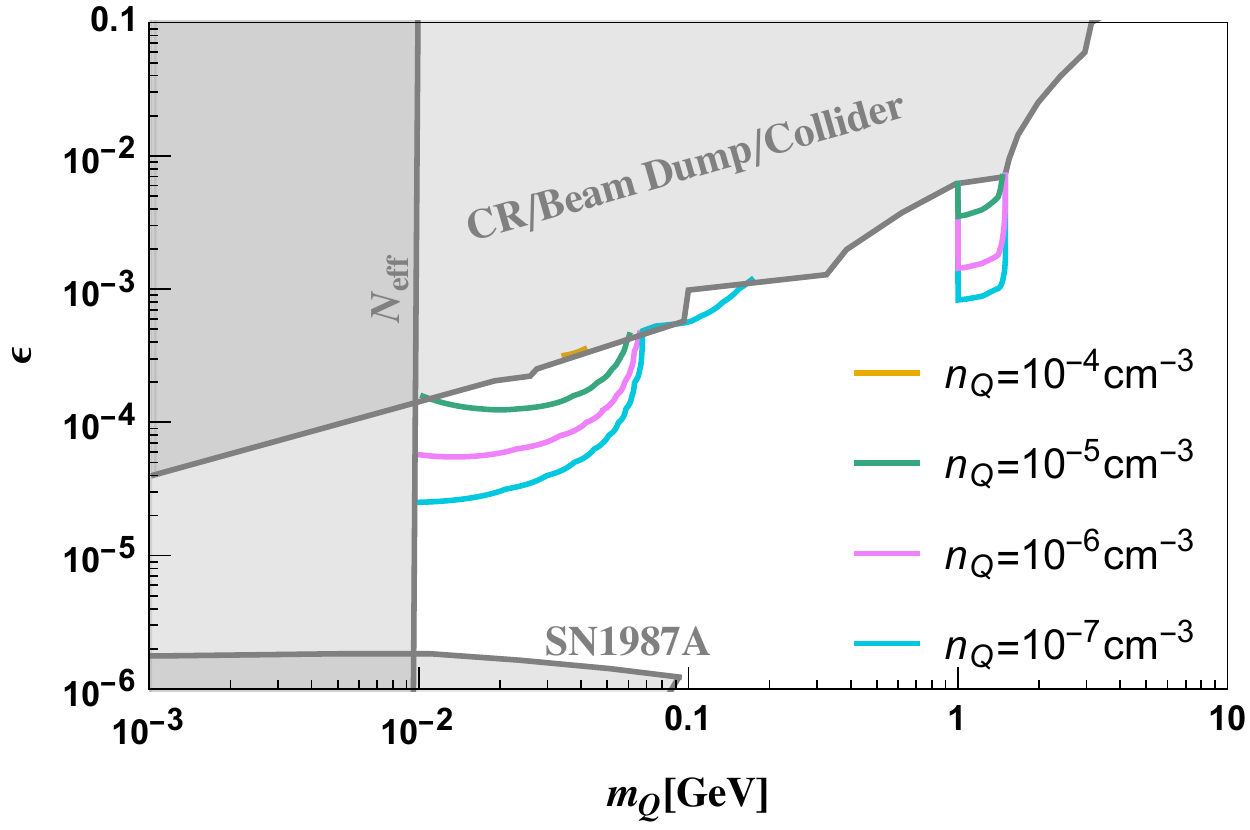} 
        \label{fig:crcap2}
    \end{subfigure}
\centering
 \caption{Accumulated terrestrial density of mCPs arising from decay of mesons produced by cosmic rays in the atmosphere. \textbf{Left}: number densities neglecting evaporation; \textbf{Right}: realistic number densities upon accounting for evaporation. Also shown are a compilation of existing constraints on mCPs from cosmic rays~\cite{Plestid:2020kdm}, beam dumps and colliders~\cite{Marocco:2020dqu}. $N_{\rm eff}$ constraints from \cite{Creque-Sarbinowski:2019mcm} are displayed as well. We also show the region that is accessible to surface and deep underground direct detection (DD)~\cite{Emken:2019tni}}
 \label{crcap}
\end{figure*}

\begin{figure*}[htpb]
\centering
  \begin{subfigure}[t]{0.49\textwidth}
        \centering
        \includegraphics[width=\linewidth]{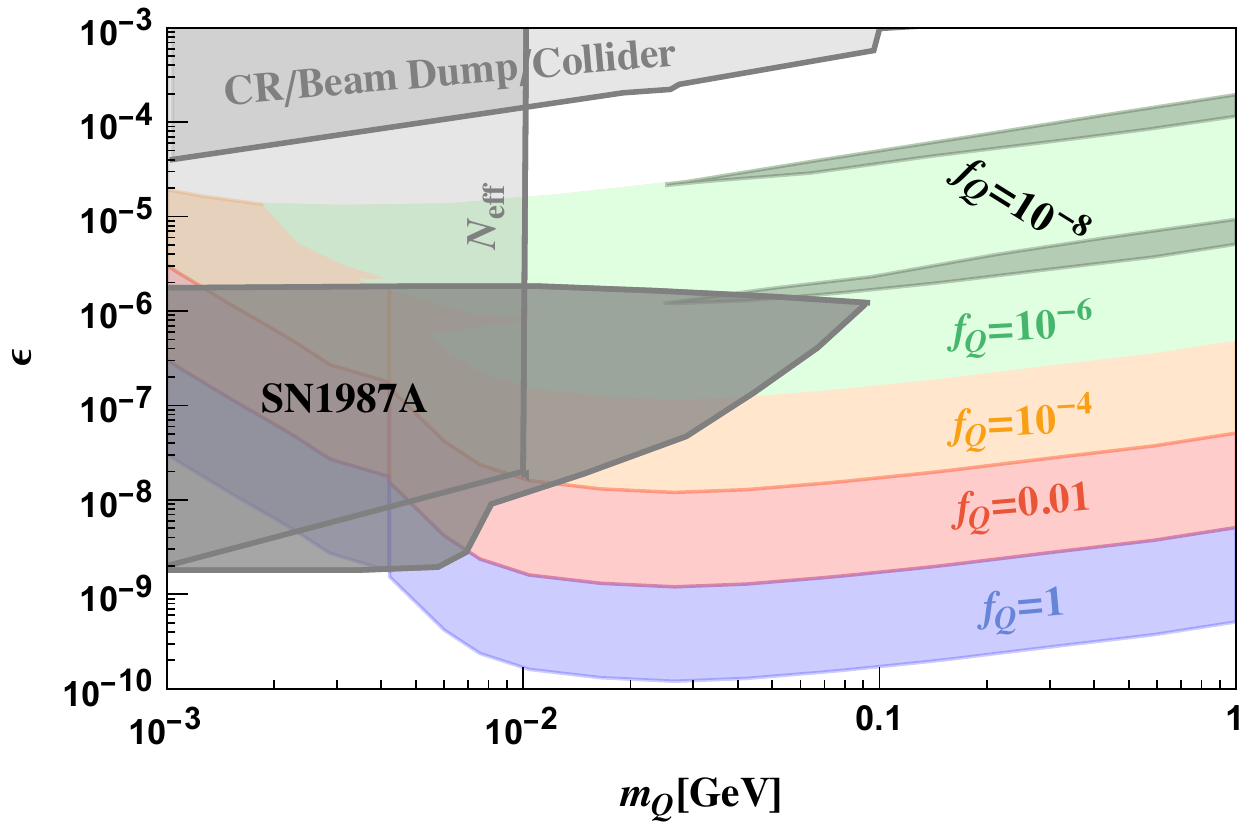} 
    \end{subfigure}
        \hfill
    \begin{subfigure}[t]{0.49\textwidth}
        \centering
        \includegraphics[width=\linewidth]{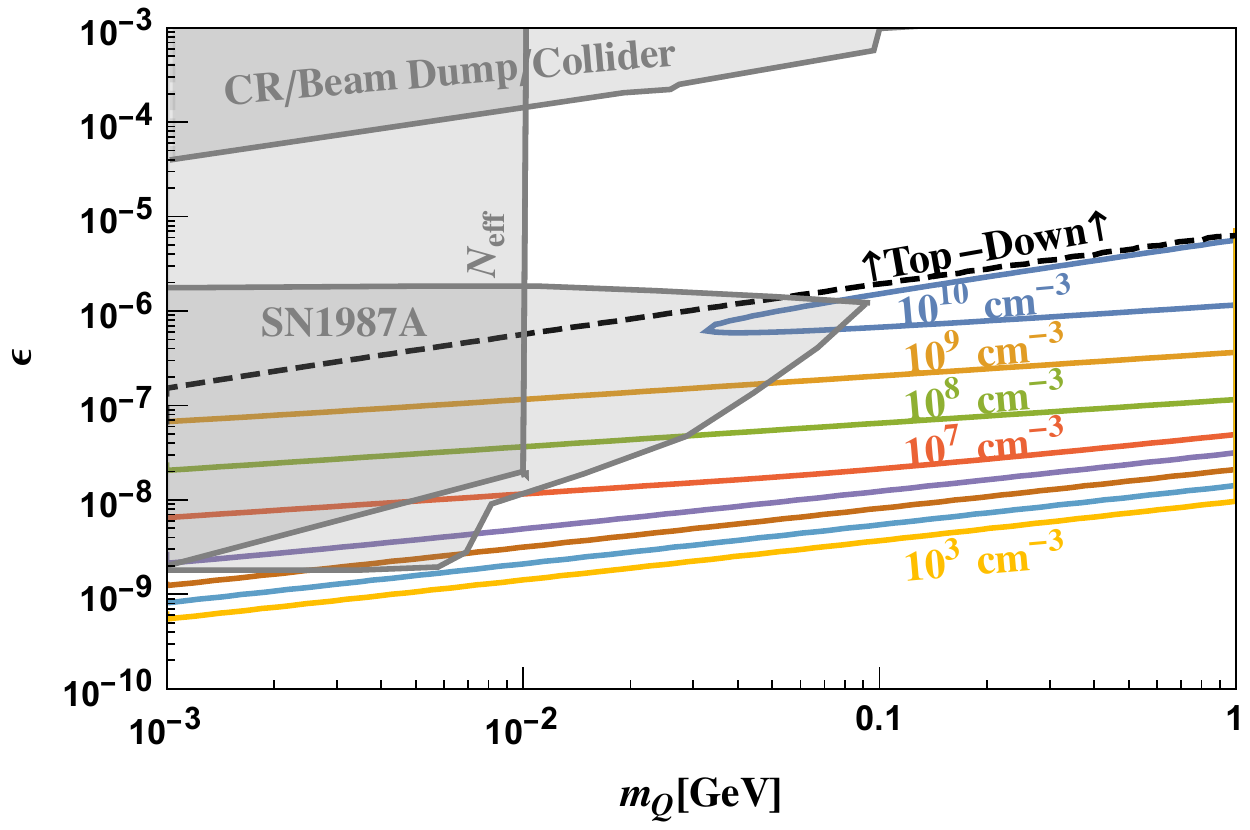} 
    \end{subfigure}

\centering
 \caption{\textbf{Left}: Existing DD limits on mCDM parameter space rescaled from the $f_Q=1$ constraints of~ \cite{Emken:2019tni}, \textbf{Right}: Contours of $n_Q/f_Q$ arising from accumulation due to virial mCDM density are plotted.}
 \label{lowchargefig}
\end{figure*}

The previous section dealt with top-down accumulation; mCPs rapidly thermalizing in the overburden followed by diffusion/gravity populating lower altitudes. 
However, mCPs with large enough rigidity ($\frac{\rm momentum}{\rm charge}$) could penetrate the overburden and get deep into the Earth before thermalizing. They then diffuse through rock and the atmosphere before finally evaporating. This diffusion time effectively acts as the time of accumulation of these mCPs leading to moderate local density. This large rigidity could arise either due to mCP possessing large momenta or small charge. An irreducible source of a fast flux occurs due to cosmic ray produced mesons which decay into mCPs. The flux for such mCPs was treated in detail in \cite{Plestid:2020kdm,Harnik:2020ugb}, and we do not repeat it here. An alternative source of the fast flux could be the cosmic ray collisions with mCDM that accelerates mCDM particles to higher velocity via Rutherford scattering \cite{Harnik:2020ugb}. Finally, virial mCPs with small enough charge $\epsilon$ could also penetrate the overburden and diffuse subsequently.

\subsection{Fast Flux}
We next estimate the accumulation of mCPs due to the atmospheric fast flux. In the absence of evaporation, and in the assumption that all mCPs generated in the atmosphere are captured and retained, we would have,
\begin{equation}
n_{\rm loc}=\int^{\beta \gamma_{\rm max}} d(\beta \gamma)  \frac{d\Phi}{d(\beta \gamma)} \frac{\pi R_{\oplus}^2 }{\frac{4}{3} \pi R_{\oplus}^3} t_{\oplus}
\label{nloccrnoevap}
\end{equation}
where $\frac{d\Phi}{d(\beta \gamma)}$ is the incoming mCP flux per interval of $\beta\gamma$. An mCP with mass $m_Q$, charge $\epsilon$ and boost factor $\beta \gamma $ penetrates a distance $d_{\rm pen}(m_Q,\epsilon,\beta \gamma )$ in the rock, that we estimate using the Bethe-Bloch formula. 
We cut off the integral approximately at $\beta \gamma_{\rm max}$, above which particles cannot be stopped by the entire column density of the Earth. $\beta \gamma_{\rm max}$ is given by equating the penetration depth to the Earth's radius: $d_{\rm pen}(m_Q,\epsilon,\beta \gamma _{\rm max})=R_{\oplus}$. We take the estimate for the atmospheric flux from \cite{Plestid:2020kdm}.
This quantity $n_{\rm loc}$ is plotted in Fig.~\ref{crcap} (left panel) as a function of $\epsilon$ and $m_Q$. Thus, this plot gives an expected average density created by cosmic rays, if evaporation can be neglected. Also shown are a compilation of existing constraints on mCPs from cosmic rays~\cite{Plestid:2020kdm}, beam dumps and colliders~\cite{Marocco:2020dqu}. $N_{\rm eff}$ constraints from \cite{Creque-Sarbinowski:2019mcm} are displayed as well. We also show the region that is accessible to surface and deep underground direct detection (DD)~\cite{Emken:2019tni}. We find that for $\epsilon \lesssim 5\times10^{-6}$, there is negligible terrestrial accumulation since the mCP interacts feebly enough to penetrate the entire Earth without thermalization. As $\epsilon$ is increased, there is also a larger flux due to preferential meson decays resulting in larger accumulation. We find that densities up to $n_Q\approx 1\textrm{cm}^{-3}$ can be achieved barring evaporation. 

It is clear that this density will be diminished due to evaporation, and the total local density will depend sensitively on the retention time. This can be thought as the time taken for the mCP to diffuse out to the surface (with subsequent evaporation determined by $m_Q$) is given by the diffusion time $t_{\rm diff}(d_{\rm pen})$ given in Eqn.~\ref{tdiff}.  We approximate the total number of mCPs collected in the infinitesimal shell with depth $d_{\rm pen}$ to have been distributed with linearly decreasing density in the shell of thickness $d_{\rm pen}$. Thus we have for the local density,
\begin{widetext}
\begin{equation}
n_{\rm loc}(h)\approx\int^{\beta \gamma_{\rm max}} d(\beta \gamma ) \frac{d\Phi}{d(\beta \gamma)} \frac{\pi R_{\oplus}^2 t_{\rm diff}(d_{\rm pen})
 }{\frac{4}{3} \pi \left(R_{\oplus}^3-(R_{\oplus}-d_{\rm pen})^3\right)}\frac{h}{d_{\rm pen}} \label{nloccr}\approx\int^{\beta \gamma_{\rm max}} d(\beta \gamma ) \frac{d\Phi}{d(\beta \gamma)}\frac{h}{v_{\rm th}\lambda}
\end{equation}
\end{widetext}
This quantity is plotted in Fig.~\ref{crcap} (right panel). The effect of evaporation is severe for lighter masses, due to their superior thermal velocities which leads to shorter diffusion times. Above a GeV, evaporation is negligible and the left and right panels present near identical densities. In the region currently allowed by terrestrial bounds, densities upto $n_Q\approx \frac{10^{-4}}{\textrm{cm}^3}$ can be achieved. While this is several orders of magnitude smaller than the densities found in Section~\ref{vdm} for mCDM, it is important to note that this is an irreducible density with no assumptions regarding the relic density of these mCPs.

\subsection{Virial mCDM with Small $\epsilon$ }

Alternatively, virial mCPs with lower charge $\epsilon$ could also reach significant depths before thermalizing. This thermalized mCP then diffuses outwards before eventually evaporating. Unlike the model variation presented in Section~\ref{vdm}, these mCPs can reach surface and underground detectors without significant slow-down, leading to strong limits from existing DD experiments shown in  Fig.~\ref{lowchargefig}, left panel, which we rescaled from the $f_Q=1$ constraints shown in Ref.~\cite{Emken:2019tni}. However, as seen in the figure, the limits relax for subcomponent mCDM, with no existing limits below $f_Q\approx 10^{-8}$. 

We next calculate the local thermalized density to explore a complementary probe of this parameter space. 

The local density can be calculated using Eqn.~\ref{nloccr} with the flux given by,
\begin{equation}
    \frac{d\Phi}{d(\beta \gamma) } \approx \frac{d\Phi}{dv_Q}=v_Q \frac{f_Q \rho_{\rm DM}}{m_Q} g(v_Q)
\end{equation}
where the non-relativistic approximation $\beta \gamma \approx v_Q$ is used, and $g(v_Q)$ is the Maxwell Boltzmann distribution boosted to the Earth frame given in Appendix B of \cite{Essig:2015cda}. 

We provide contours of the resultant mCDM density at 1km depth in rock in Fig.~\ref{lowchargefig} (Right panel). For small enough $\epsilon$ the entire Earth is not enough to stop virial DM and hence there is no accumulation. Above the black dashed line, mCDM stops in the overburden and we leave the estimation of the number densities for top-down accumulation to future work. 

We note that for part of this parameter space
({\em e.g.} for $\epsilon\sim 10^{-6}$, $m_Q\sim 100\,{\rm MeV}-1\,{\rm GeV}$ and $f_Q \sim 10^{-8}$ concentrations of mCPs in the underground laboratories can reach $10^2$\,cm$^{-3}$, which may still provide some basis for future detection, despite the smallness of $\epsilon$, as discussed in Section VI.

\section{Bound states}
\label{boundstates}
The thermalized dark matter now has kinetic energy on the order of 
$kT_{\rm room} \sim 0.025~\textrm{eV}$. The negatively charged mCDM can now form bound states with atoms. At a sufficiently large mass, the lowest orbit of such bound states is inside the atomic K-shell if, 

\begin{align}
r&=\frac{a_0}{Z \epsilon} \frac{m_e}{\mu_{Q,N}} < \frac{a_0}{Z }  \nonumber \\
\epsilon &> \frac{m_e}{\mu_{Q,N}}
\label{negineq}
\end{align}
Here $\mu_{Q,N}$ is the reduced mass of the mCDM-nuclear system, $a_0$ is the Bohr radius, and $Z$ is the atomic number of the nucleus. If this condition is satisfied, the binding energy will be on the order of $E_B \simeq Z^2 \epsilon^2 \mu_{Q,N}/2\simeq
13.6\,{\rm eV} \times Z^2\epsilon^2(\mu_{Q,N}/m_e)$.

One may worry that if the bound states form, the existing atomic electron gains some positive energy due to effective screening of the atomic nucleus, $Z\to Z-\epsilon$.  The total binding energy of an atom scales as $16 \textrm{eV} Z^\frac{7}{3}$. This is obtained by observing that to a good approximation, the electrons $Z_e=Z$ in number at a distance $a_0Z_N^{-\frac{1}{3}}$ from the nucleus with effectively un-screened charge $Z_N\simeq Z$. Substituting $Z_N\rightarrow Z_N - \epsilon$, the net binding energy is

\begin{align}
\Delta E_B=  \left(13.6 \epsilon^2 Z^2 \frac{\mu_{Q,N}}{m_e}-21.3 \epsilon Z^\frac{4}{3}\right)  \textrm{eV}
\end{align}
Requiring $E_B>0$ gives
\begin{align}
\epsilon > \frac{8\times 10^{-4} }{Z^\frac{2}{3} }\frac{\rm GeV}{\mu_{Q,N}}.
\end{align}
This is always weaker than Eqn.~(\ref{negineq}) for $Z>1$, and therefore we take (\ref{negineq}) as the main criterion for the bound state formation by negatively charged mCPs. 

These exotic bound objects have net charge $ \epsilon$. The nucleus could attract more negative charged mCPs but it is unlikely because there are fewer mCPs than nuclei. They could form bound states with positively charged mCPs as well. 

The positive mCPs could form bound states with free electrons when the bound state energy exceeds their kinetic energy due to thermal equilibrium with the room temperature,
\begin{align}
E_B&=13.6~\textrm{eV} \epsilon^2  > 0.025 \textrm{eV}  \nonumber \\
\implies \epsilon &> 0.042.
\label{Qe}
\end{align}
or could form bound states with negatively charged mCDM / atom-negative charge hybrids. This bound state has energy,
\begin{align}
E&=13.6~\textrm{eV} \epsilon^4 \frac{m_Q}{m_e}  > 0.025 \textrm{eV}  \nonumber \\
\implies \epsilon &> 0.2 \left( \frac{m_e}{m_Q} \right)^\frac{1}{4}
\label{QQ}
\end{align}
Notice, however, that the bound states of $Q^+Q^-$ can occur due to $A'$ exchange, and can be significantly deeper than electrostatic bound states of the same particles. The existence of the $Q^+Q^-$ binding can have profound consequences for cosmological abundances and late annihilation of mCP, as is discussed {\em e.g.} in Refs. \cite{ArkaniHamed:2008qn,Pospelov:2008jd,An:2016gad,Cirelli:2016rnw}. 

\begin{figure}[htpb]
\centering
\includegraphics[width=0.48\textwidth]{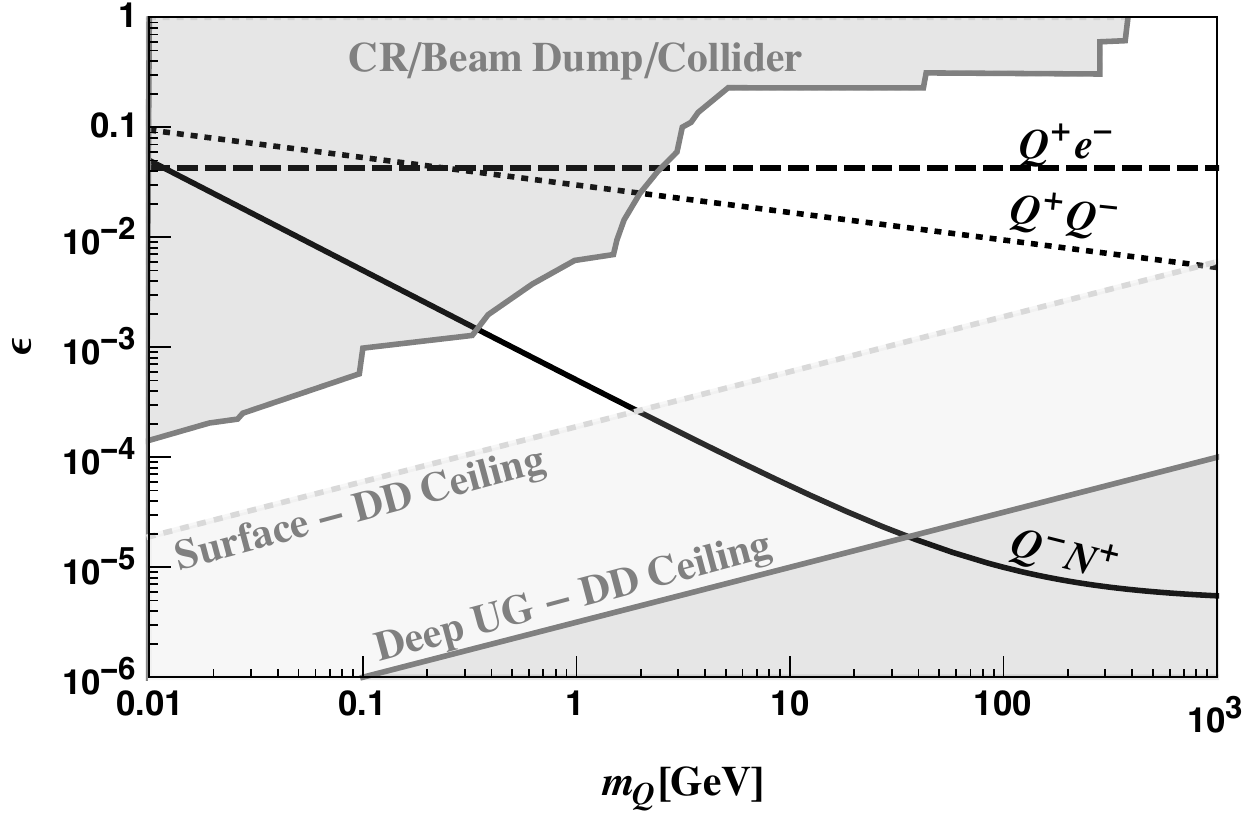}
\caption{Parameter space that allows for the bound states of mCPs with nuclei (Solid), electrons (Dashed) and between themselves (Dotted) at $300$ K. The line corresponds to the bound state energy $E_B=300$~K and the region above corresponds to deeper bound states.} 
\label{fig4}
\end{figure}

In both Eqs. (\ref{Qe}) and (\ref{QQ}), we have required that the depth of these bound states exceed typical thermal energy, because otherwise they can be easily broken up by thermal collisions. 
An approximate position of the critical dividing lines for the bound states is shown in Fig. ~\ref{fig4}. The region above the lines correspond allow stable bound states at $300$~K. As is easy to understand, the electrostatic attraction between $Q^-$ and a nucleus is strongest, and the solid black line is plotted for a typical nucleus with $Z\sim 30$.

We briefly outline observable physics effects that can occur due to a formation of $(Q^-N^+)$ bound states. The cross section leading to these bound states is not necessarily small. A typical formation of this bound state will occur with an Auger-type ejection of an electron, and subsequent cascade of the bound state down to its ground state,
\begin{equation}
\label{recomb}
    {\rm Atom}+Q^-\to ({\rm Atom}^{+n}Q^-)+n e+(\gamma).
\end{equation}
The rate for such a process can be large, and we make a crude estimate of the cross section by accounting for the relatively small size of the nucleus-$Q^-$ bound state, $\sim \pi (a_0/Z)^2$, the probability of an outer electron to be within that distance from the nucleus $\sim Z^{-2}$. The cross section will contain $1/v_Q$, the inverse velocity of the incoming particle, which will be made dimensionless by the typical velocity of an electron inside the $K$-shell, $\sim Z\alpha$. This way, one get the following estimate for the cross section of bound state formation, 
\begin{equation}
\label{capture}
\sigma_{\rm capture} v_Q \sim \frac{\pi \alpha a_0^2\times c}{Z^3} \propto 10^{-23}\,{\rm cm}^2\times c \times (30/Z)^3.
\end{equation}
This size of the cross section will ensure relatively rapid capture of $Q^-$, if the bound state formation is possible. The refinement of this estimate along the line of computations performed in \cite{Pospelov:2008qx,An:2012bs} is possible. 

The capture of $Q^-$ by the nuclei of light elements may lead to exotic concentrations of $({\rm H}Q)$, $({\rm C}Q)$ and $({\rm O}Q)$. However, the search techniques that involve ionization and mass spectrometry \cite{Smith:1982qu,Hemmick:1989ns}, as well as ``alternative" chemical history for the milli-charged bound states poses certain difficulties in applying such bounds. 

A less uncertain approach to search for an mCP ``recombination" with an atom would consist in searching for heat/ionization provided by the process (\ref{recomb}). For example, just below the $(Q^-N^+)$ boundary in Fig. ~\ref{fig4} the recombination with light elements is not possible but recombination with atoms such as Xe or I happens readily. An ideal setup for such a probe would be the DM-Ice experiment \cite{deSouza:2016fxg} that utilizes NaI crystals shielded by $\sim$2.5 km of ice. Assuming the range of parameters that does not allow the formation of bound states with H, O and elements in the atmosphere, one could still expect - for a right range of $\{m_Q,\epsilon\}$, the exothermic reaction of $Q^-$ association with iodine atoms. 
Taking into account that the capture cross sections can be significant (\ref{capture}), all (or nearly all) of the negative mCP incident on NaI crystal may undergo the capture process. In this case one should expect that the counting rate is 
\begin{eqnarray}
\frac{\rm Events}{\rm time} &\sim n_Q(2.5\,{\rm km})\times v_{\rm th} \times {\rm Area} \nonumber\\
&\propto 10^{6}{\rm Hz} \times \frac{n_Q}{1\,{\rm cm^3}} \times \frac{\rm Area}{100\,\rm cm^2}\times \left(\frac{100\,\rm GeV}{m_Q}\right)^{1/2}.
\end{eqnarray}
If the binding energy is between few keV to a 100 keV, 
one should compare it with the counting rates observed by these experiments that do not exceed $O(10)\, {\rm kg}^{-1}{\rm day}^{-1} {\rm keV}^{-1}$, which for 10 kg crystals and two decades in energy does not exceed the total counting rate of $0.1$\,Hz. Therefore, for 100\,GeV particles it translates to sensitivity to $n_Q$ at the level of $10^{-7}\,{\rm cm}^{-3}$, and given results of Fig. 1, to $f_Q$ as small as $10^{-19}$. Conversely, one can achieve some sensitivity to cosmic ray generated mCP flux. Further gains in sensitivity can be achieved by exploiting that one and the same amount of energy is released in the formation of the bound states with a given atom, which would then show as an ``unidentified" line in the spectrum taken by a detector.

We finish this subsection by acknowledging the fact that the full exploration of the sensitivity to bound states throughout $\{m_Q,\epsilon\}$ parameter space is difficult, as the binding to lighter elements will change patterns of mCP accumulation and distribution with depth. We leave more detailed exploration of the bound state related observables to forthcoming work.

\section{Annihilation inside large volume detectors}
\label{annihilations}
\begin{figure*}[htpb]
\centering
  \begin{subfigure}[t]{0.49\textwidth}
        \centering
        \includegraphics[width=\linewidth]{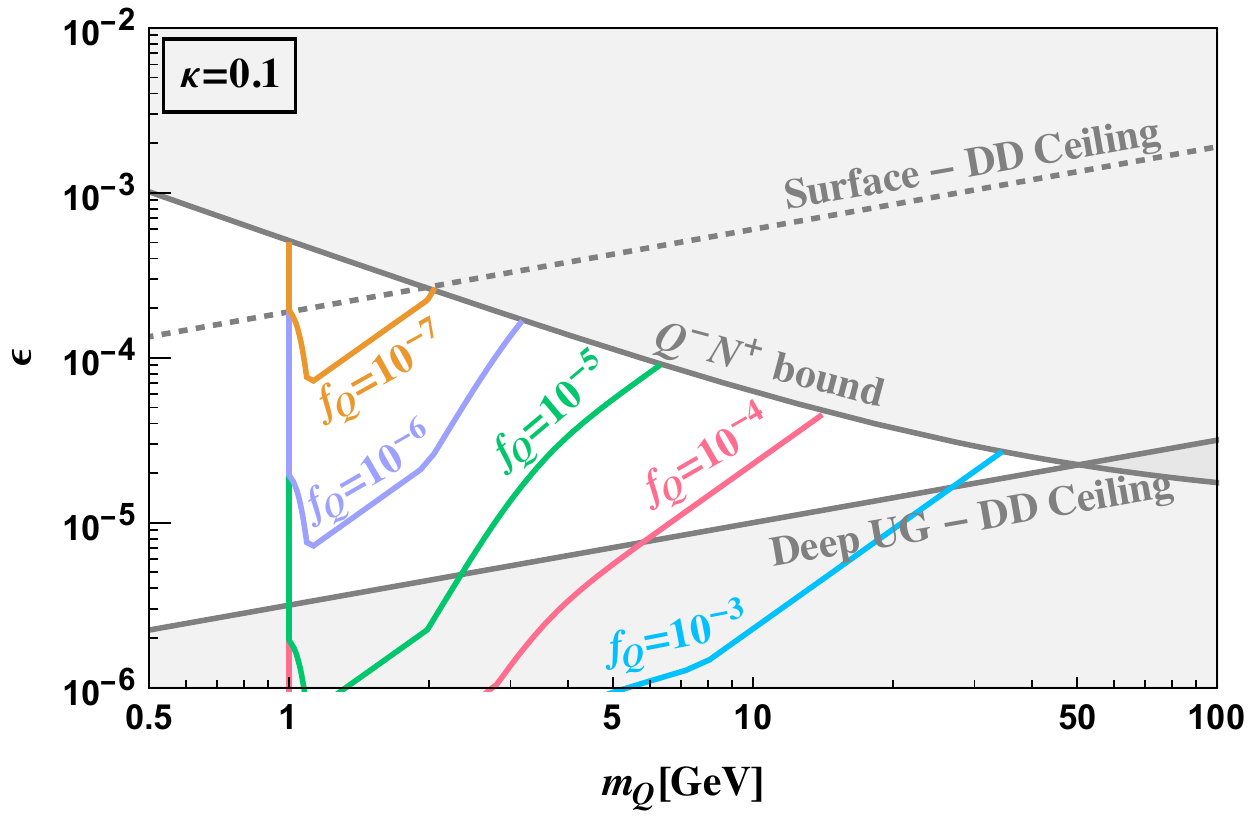} 
    \end{subfigure}
        \hfill
    \begin{subfigure}[t]{0.49\textwidth}
        \centering
        \includegraphics[width=\linewidth]{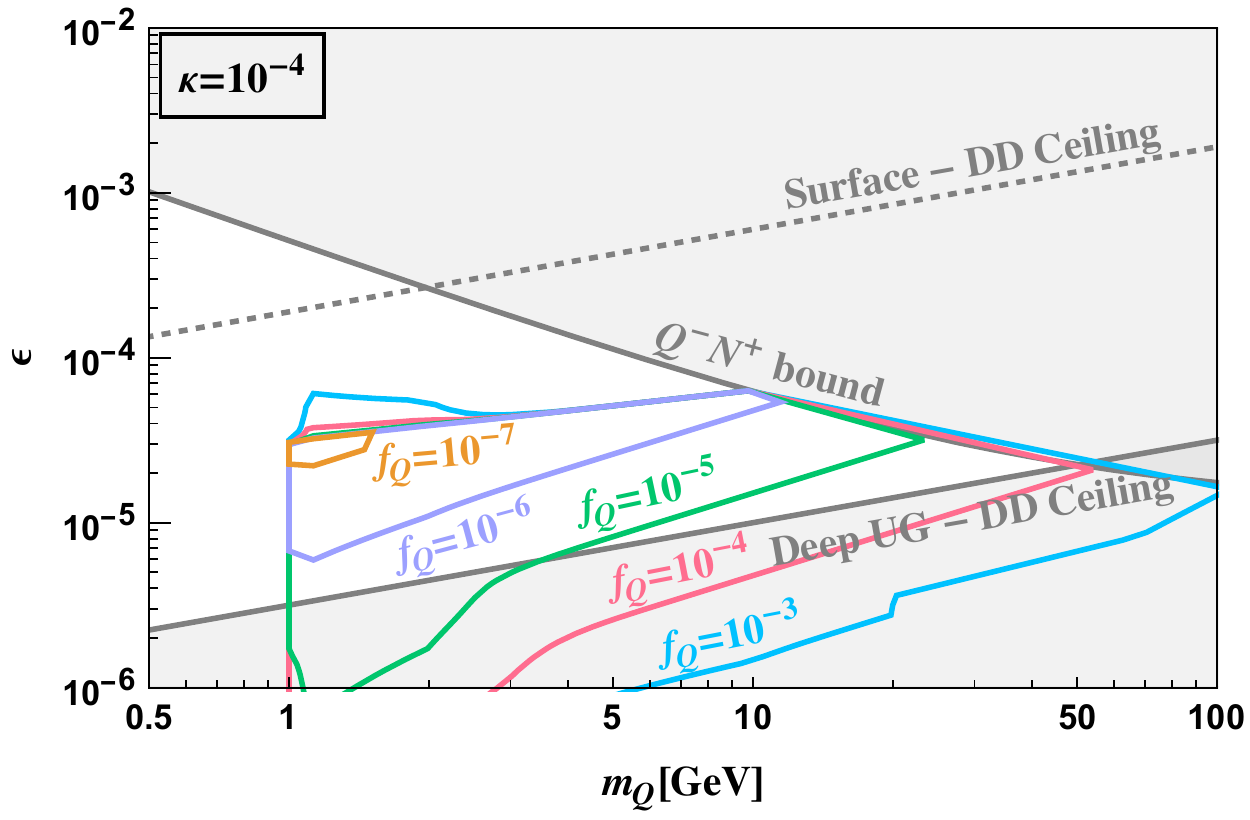} 
    \end{subfigure}

\centering
 \caption{Limits corresponding to more than 200 events in 22 kton year exposure at Super-K for $Q\bar{Q}\rightarrow e^+ e^-$ for different DM fractions $f_Q$ for two different choices of the kinetic mixing parameter $\kappa$, \textbf{Left}: $\kappa=0.1$ and \textbf{Right}: $\kappa=10^{-4}$ are shown. For the same millicharge $\epsilon$, smaller $\kappa$ corresponds to larger $\alpha_D$ and hence larger Sommerfeld enhancement for annihilation. This leads to limits shifting downward for smaller $\kappa$.}
 \label{superk}
\end{figure*}
In this section we explore the possibility of mCDM being in equal amounts of particle anti-particle pairs. The presence of a large number of positive and negative mCDM terrestrially can lead to annihilations. Notice that the negative charge can be bound deep inside atoms, and the probability of annihilation may get significantly reduced due to the electrostatic repulsion. (A positively charged mCP would not be able to approach the orbit of bound negatively charged mCP). On the other hand, the $A'$-induced attraction between the two mCP particles may actually overcome the Coulomb repulsion, and the annihilation may proceed even if the negative mCP is locked inside an atomic bound state. 
Unfortunately, reliably predicting the abundance of mCP at locations of underground laboratories when negative mCPs are intercepted by atoms is extremely challenging. 
For this reason, we concentrate on the region of parameter space where both the positive and negative mCDM are unbound by atoms. 

Annihilation of mCPs can occur via a variety of different channels. The largest cross section presumably occurs 
due to annihilation to two dark photons, $Q \bar Q \to A' A'$. The cross-section for this process, including Sommerfeld enhancement for small velocities is \cite{ArkaniHamed:2008qn},
\begin{equation}
\sigma_{\rm ann}^{D} v_Q=\frac{\pi \alpha_D^2}{m_Q^2} \frac{\pi \alpha_D}{v_Q\left(1-e^{-\pi \alpha_D/v_Q}\right)}
\end{equation}
Here, $\alpha_D=\frac{g_D^2}{4\pi}$ and the superscript $D$ denotes dark final state to differentiate from annihilations to SM which we will look at shortly. $v_Q$ is the velocity of each particle which we will take to be the thermal velocity $v_{\rm th}$. This large annihilation rate can result in depletion of the large densities calculated in Section.~\ref{vdm}. The Jean's density $n_{\rm jeans}$ calculated in \cite{Neufeld:2018slx} is a result of accumulation through the age of the Earth, and we find that with annihilations turned on, it is
\begin{align}
n_{\rm jeans}^{\rm ann} = \frac{\tanh(\zeta)}{\zeta} n_{\rm jeans}
\end{align}
Here $\zeta^2=\frac{3n_{\rm vir}v_{\rm vir} t_{\oplus}^2 \sigma_{\rm ann}^{D} v_{\rm th}  }{4R_E}$. Note that for $\sigma_{\rm ann}^{D}\rightarrow 0$, $\frac{\tanh(\zeta)}{\zeta}\rightarrow 1$ as expected. In the opposite limit, i.e. for large cross-sections, $\frac{\tanh(\zeta)}{\zeta}\rightarrow \frac{1}{\zeta}$ . 

The traffic jam contribution takes on average a time duration, $\frac{L_{\rm ob}}{v_{\rm sink}} \ll t_{odot} $ before reaching the detector and hence the suppression is smaller. Here $L_{\rm ob}\approx 1~\textrm{km}$, is the length of the overburden. We conservatively estimate the traffic jam density to be
\begin{align}
n_{\rm tj}^{\rm ann}=0.2\frac{v_{\rm term}}{L_{\rm ob}  \sigma_{\rm ann}^{D} v_{\rm th}}
\end{align}
i.e. $0.2$ of the density that corresponds to the DM column density required to annihilate all of the incoming thermalized DM.
Finally, we use 
\begin{equation}
n_{\rm loc}^{\rm ann} = \textrm{Max}\left( n_{\rm jeans}^{\rm ann}, \textrm{Min}\left( n_{\rm tj}^{ann} , n_{\rm tj},  \langle n_Q \rangle \right)\right)
\label{locann}
\end{equation}
Thus, the local density of mCPs in the lab is critically dependent on the dark fine structure $\alpha_D$, with larger  $\alpha_D$ leading to smaller local abundances. 
We next turn to observables produced by $Q^+Q^-$ annihilation. While annihilation to two $A'$ is dominant, and subsequent conversion/oscillation of $A'$ into $A$ can occur, it is severely suppressed at small mass of $A'$. Hence we look at annihilation into SM in this work.

{\em Visible} annihilation channels via an $s$-channel virtual photon, $Q \bar Q \to A^*\to {\rm SM} $, will occur at a rate suppressed by $\epsilon$ but will 
result in immediate release of energy in the form of SM charged particles.

The annihilation cross-section of $Q\bar{Q}$ into $e^+e^-$ is given by,
\begin{align}
    \sigma_{\rm ann}^{\rm SM} v&=\frac{\pi\epsilon^2\alpha^2 }{3m_Q^2}  \frac{\pi \alpha_D}{v_Q\left(1-e^{-\pi \alpha_D/v_Q}\right)} 
\end{align}

Next, let us estimate the number of annihilation events inside a volume that we will take to correspond to the fiducial volume $V$ of the Super-Kamiokande experiment. The event rate is given by,
\begin{align}
    (n_{\rm loc}^{\rm ann})^2 \sigma_{\rm ann}^{\rm SM} v_{\rm th} V = 400 \left(\frac{\epsilon}{10^{-5}}\right)^2 \left(\frac{n_Q}{10^{8}/\textrm{cm}^3}\right)^2 \\ \nonumber \times \frac{V}{{22000\,\rm m}^3} \left(\frac{\rm GeV}{m_Q}\right)^2 \frac{1}{\textrm{year}} \times \{4\times10^{5}\alpha_D\sqrt{m_Q/\textrm{GeV}}\}.
 \end{align}
 which potentially may result in a very strong sensitivity to $\epsilon \times n_Q m_Q^{-1}$. The quantity between the $\{\}$ provides the additional Sommerfeld enhancement. Sensitivity to $\alpha_D$  enters through this enhancement factor as well as through $n_{\rm loc}^{\rm ann}$ in Eqn.~\ref{locann}.

Fig.~\ref{superk} illustrates this through limit contours for the different input values of $f_Q$. We take $\mathcal{O}(200)$ events per year to be roughly a limiting count rate. The left panel corresponds to $\kappa=0.1$ (small $\alpha_D$) and the right panel corresponds to $\kappa=10^{-4}$ (large $\alpha_D$). For a fixed small $\epsilon$, larger $\alpha_D$ results in larger Sommerfeld enhancement and thus stricter limits on $f_Q$. However, sensitivity to large $\epsilon$ is lost for large $\alpha_D$ because this results in high annihilation rates as well, and hence extremely small local number densities $n_{\rm loc}^{\rm ann}$. Nonetheless new parameter space is probed for both the small $\alpha_D$ and large $\alpha_D$ regimes.


\section{Electrostatic accelerators}
\label{accelerator}

\begin{figure*}[htpb]
\centering
\includegraphics[width=\linewidth]{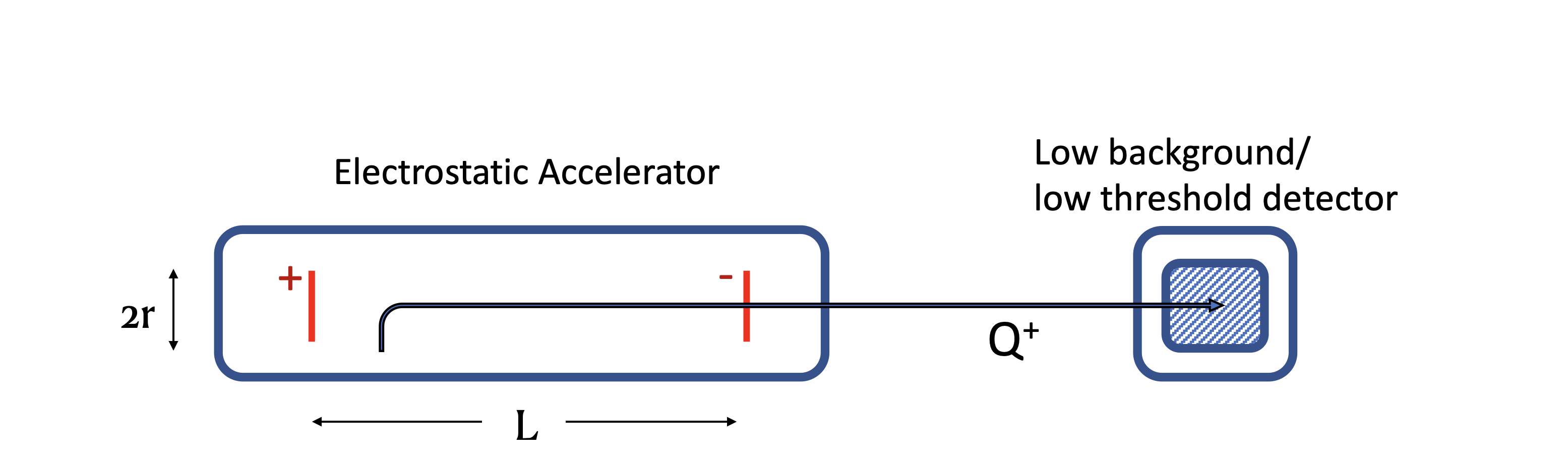} 
\caption{Scheme of electrostatic accelerator concept. Positive mCPs between the two plates held at a potential difference $\Delta V$ get accelerated towards the more negatively charged plate and enter the low threshold detector to be detected.}
\label{scheme}
\end{figure*}

Self-annihilation, enhanced by terrestrial accumulation, provides non-trivial limits on the parameter space of the mCDM model. At the same time, quadratic scaling with abundance does not allow to probe very small $f_Q$. The smallest $f_Q$ where experiments like Super-K will have sensitivity is for $f_Q\sim 10^{-7}$. In this section we propose a novel strategy to test even smaller densities. 

As alluded to in the Introduction, the local density of mCPs could be accelerated in a large electric field and the accelerated mCPs could then be detected. Owing to disparate charge to mass ratio compared to SM particles, oscillating field accelerators will not be suitable for mCPs. Instead electrostatic accelerators such as Van de Graaf generators and Cockcroft-Walton accelerators would be suitable. Modern accelerators with potential difference ($\Delta V$) in the megavolt range are used in nuclear physics experiments. Examples among these are LUNA ($\Delta V=3.5$ MV)  \cite{sen2019high}, JUNA ($\Delta V=0.4$ MV) \cite{liu2016progress} and CASPAR  ($\Delta V=1.1$ MV) \cite{Robertson:2016llv}.

Since we are considering mCPs with a dark photon, we need to first determine the plasma masses so as to ensure that the electric field is not shielded by mCPs. The plasma mass of the dark photon $A'$ in the presence of a number density $n_Q$ of mCPs $Q$ is given by, 
\begin{equation}
    \Pi=g_Q^2 \frac{n_Q}{m_Q}\approx (3\times 10^{-5} \textrm{eV})^2 \frac{n_Q}{10^{14}/\textrm{cm}^3} \frac{\textrm{GeV}}{m_Q}
    \end{equation}
Here $g_Q\rightarrow 1$ is set, to be conservative. The range of the electric field is thus,
\begin{equation}
    \lambda \approx 7 \textrm{mm} \times  \sqrt{\frac{10^{14}/\textrm{cm}^3} {n_Q}}\sqrt{\frac{m_Q}{\textrm{GeV}}}
    \end{equation}
   In other words, the screening length is larger than 1 meter for $n_Q \lesssim \frac{10^8}{\textrm{cm}^3}$. If the concentrations of mCPs exceed this level, it is likely that even the acceleration of ``normal" protons will get compromised.

    We consider the  accelerator field to be turned on, but with the proton source inside the accelerator being ``off". While the mCPs outside the region with the electric field receive no net acceleration, mCP particles on the inside  may get accelerated. Given a $E_{\rm thr}$ required for detection, it is clear that in order to have sensitivity one should require
\begin{equation}
\epsilon e \Delta V > E_{\rm thr},
\end{equation}
where $\Delta V$ is the accelerating voltage. 

To create a more realistic description, we model the accelerator tube as a $r=1$ mm radius, $L=1$ meter long tube similar to the LUNA setup  \cite{sen2019high} at 1 km depth. This concept is illustrated schematically in Fig.~\ref{scheme}. 
The flux of mCPs seeping into the pipe that gets accelerated to detectable energies is given by
\begin{align}
\Phi[E>E_{\rm thr}]&=2\pi r L\left(1-\frac{E_{\rm thr}}{\epsilon e \Delta V} \right)  \frac{f_Q \eta \rho_Q}{m_Q}  v_{\rm th}\nonumber \\ &\times \textrm{Min}[1,\frac{r \epsilon e \Delta V}{L \sqrt{T E_{\rm thr}}}] \nonumber \\
&=2\times 10^{21}  \textrm{Hz} f_Q \left(\frac{\rm GeV}{m_Q}\right)^\frac{5}{2}\left(1-\frac{E_{\rm thr}}{\epsilon e \Delta V} \right)\nonumber \\ &\times  \textrm{Min}[1,\frac{r \epsilon e \Delta V}{L \sqrt{T E_{\rm thr}}}] 
\end{align}
This expression is derived in detail in Appendix.~\ref{accgeom}. 

In Fig.~\ref{rateinluna}, we plot the rate for accelerated mCPs to come out of this setup. The accelerated flux of the mCPs is relatively collimated, and could be detected with relatively compact dark matter detectors. To translate the rate of accelerated mCPs to the counting rate we need to estimate the probability of generating one signal event by a single accelerated mCP particle.

\begin{figure*}[htpb]
\centering
  \begin{subfigure}[t]{0.48\textwidth}
        \centering
      \includegraphics[width=\textwidth]{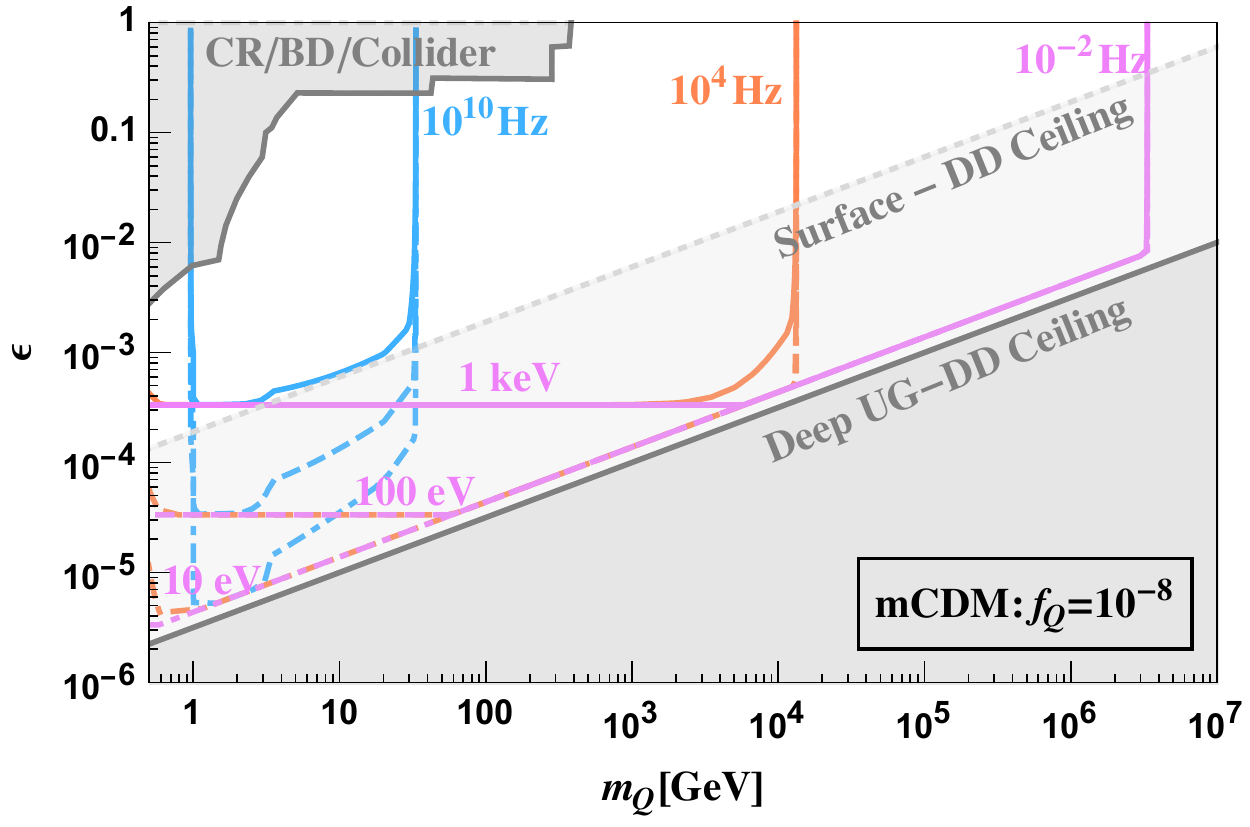}
    \end{subfigure}
        \hfill
    \begin{subfigure}[t]{0.48\textwidth}
        \centering
        \includegraphics[width=\linewidth]{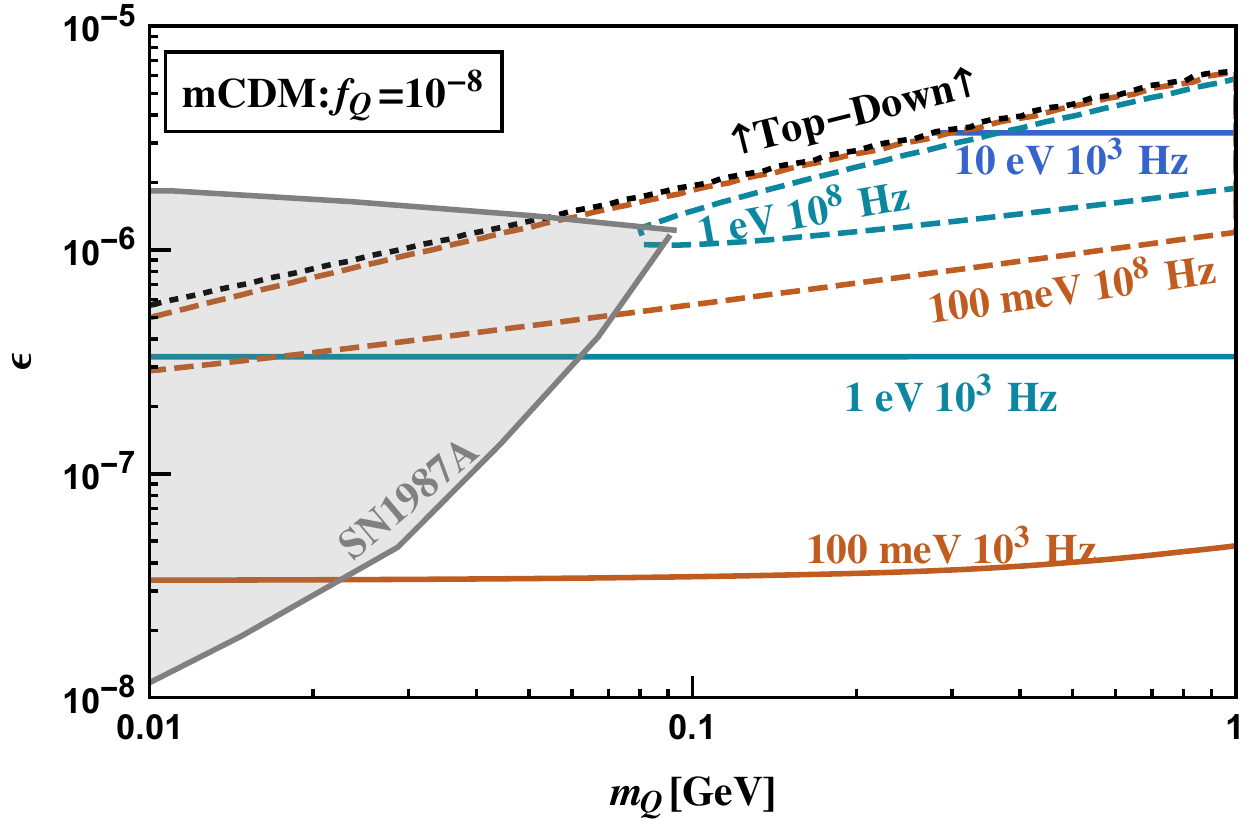} 
    \end{subfigure}

\centering
 \caption{The rates of accelerated mCDM for $f_Q=10^{-8}$ in an electrostatic accelerator at $1$~km depth. In the Left panel, the rate corresponding to $m_Q \gtrsim1$ GeV, where top-down accumulation is dominant is shown with the requirement for the final energy to exceed 10\,eV, 100\,eV, 1\,keV. In the Right panel, the rate corresponding to $m_Q \lesssim1$ GeV, where bottom-up accumulation is dominant is shown with the requirement for the final energy to exceed 100\,meV, 1\,eV, 10\,eV.} 
 \label{rateinluna}
\end{figure*}

We assume a dark matter detector that completely covers the geometric parameters of the mCP beam, but with an opening in the shielding that allows mCPs to enter. The probability for mCP to create a signal event  can be 
estimated as 
\begin{equation}
P \sim \sigma_{Q,\rm atom} \times n_{\rm atoms} \ell, ~~{\rm if\,}  P<1.
\end{equation}
where $\ell$ is taken to be $\sim 5$\,cm is a realistic size for a small-to-medium size DM detector, $n_{atoms}$ is the number density of active targets in the detector. We take $n_{\rm atoms} \simeq 4.5\times 10^{22}$\,cm$^{-3}$ to correspond to the number density of germanium atoms in Ge crystal. Finally, $\sigma_{Q,\rm atom}$ is the cross section leading to atomic recoil or ionization by the accelerated mCPs. Depending on the mass of the mCP, and the type of detector, 
the relevant scattering is on the atomic electrons or elastic scattering on a whole atom, with subsequent ionization created in the inter-atomic collisions. In the latter case, the resulting ionization energy is typically quenched by a factor of $\sim0.1$. The transfer of energy in elastic scattering on an atom is the
most efficient if $m_Q \sim m_{\rm atom}$. Using a non-pertubative estimate (Appendix A) for such a cross section, $\sigma \propto 4 \pi \times (\mu v_Q)^{-2}\propto 10^{-23} {\rm cm}^2$ for a typical mass of mCP in the 100\,GeV range, and its kinetic energy in the $\sim$ keV range, it is easy to see that for this size of the cross section, the probability of scattering within the 5\,cm detector is $O(1)$. In addition, in order to maximize the signal from such detectors in terms  of $m_Q$, it would be advantageous to use devices with a wide range of atomic masses, such as CaWO$_4$ of CRESST \cite{Abdelhameed:2019hmk}. 

The lowest thresholds for detection, and sensitivity to the lowest $m_Q$ can be achieved through scattering on electrons/atomic ionization. Despite the fact that even after the acceleration, the mCPs are relatively slow, one could apply perturbation theory for estimating the cross sections leading to ionization. Taking the outer shell atomic electron to be localized within space region $\sim a_0$, its interaction with an incoming mCP to scale as $U \sim \alpha\epsilon a_0^{-1}$, 
perturbation theory is valid as long as $U\times \Delta t \sim \alpha\epsilon v_Q^{-1} <1$ \cite{Landau1981Quantum}. Taking $m_Q$ below a few GeV, $\epsilon$ in the $10^{-6}-10^{-3}$ range and $v_Q \sim (|\epsilon e\Delta V|/m_Q)^{1/2} \sim \epsilon^{1/2} \times (10^{-2}-10^{-1})$, we see that the perturbativity condition is satisfied, and the cross section can be estimated to scale as $\pi a_0^2 \times (\alpha\epsilon v_Q^{-1})^2 $. Again, this estimate {\em exceeds} $1/(n_{\rm atom}\ell)$ making $P\sim O(1)$. Thus we conclude that the rate of accelerated mCPs in Fig. ~\ref{rateinluna} can indeed translate to a similar counting rate in a small-to-medium size detector intercepting the ``beam" of mCPs.  
\begin{figure}[htpb]
\centering
\includegraphics[width=0.48\textwidth]{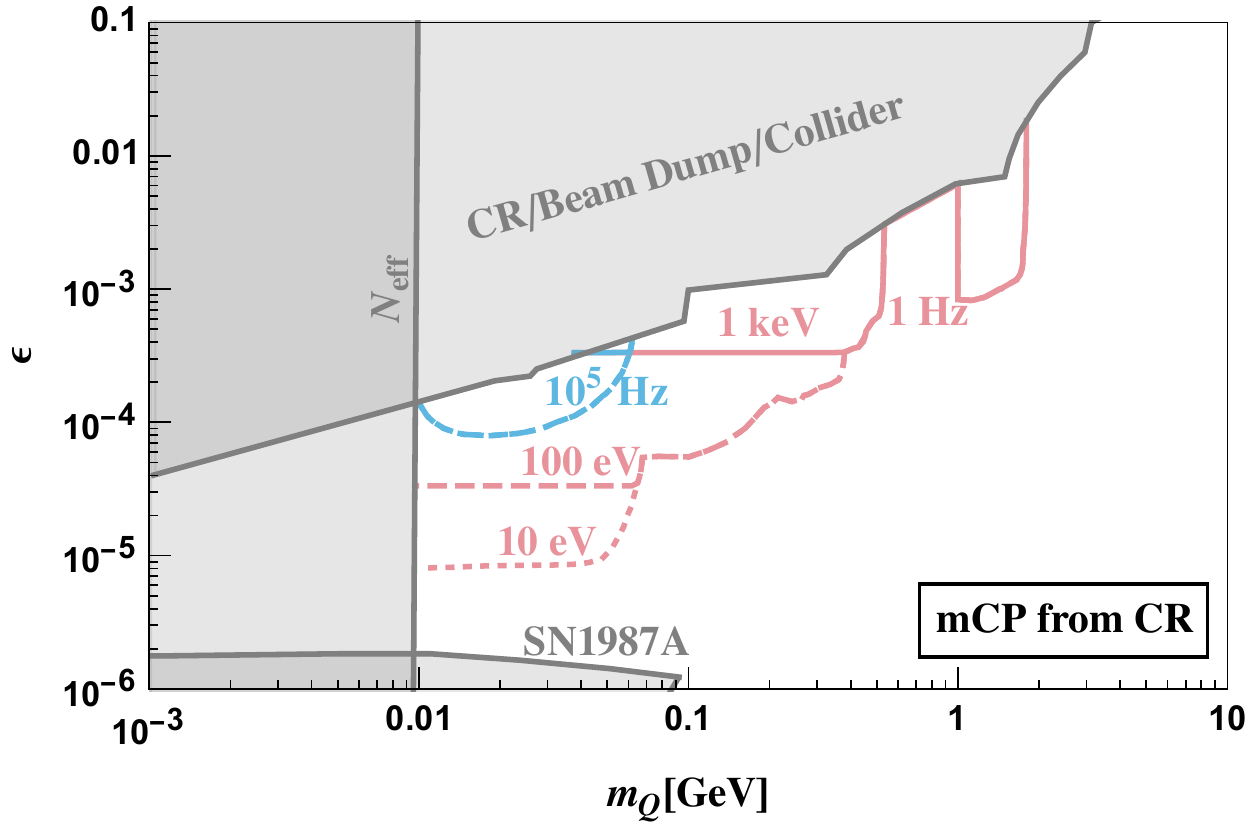}
\caption{The rates of accelerated mCPs produced in meson decays in the atmosphere in an electrostatic accelerator are shown with the requirement for the final energy to exceed 10\,eV, 100\,eV, 1\,keV }
\label{ratelunacr}
\end{figure}

Fig.~\ref{rateinluna} summarizes possible counting rates of accelerated mCPs for $f_Q=10^{-8}$. Since the probability of detection can indeed be $P\sim O(1)$, 
these rates translate to possible counting rates inside a detector placed along the path of the accelerated particles. Given that it is realistic to find detectors with background counts as low as $10^{-3}$\,Hz, it is clear that a dedicated search along the lines suggested in this section could probe $f_Q$
down to extremely small values. For $m_Q\gtrsim1$\,GeV, Fig.~\ref{rateinluna} (Left), achieving sensitivity to 
$f_Q$ as small as $10^{-20}$ looks realistic, with $f_Q\approx 10^{-8}$ achievable all the way to very large $m_Q$. 
(At large $m_Q\gg100$\,GeV one should be 
cognizant of the fact that the recoil energy of atoms 
drops as $m_{\rm atom}/m_Q$, and ionization of atoms may be suppressed if $\alpha\epsilon v_Q^{-1}$ becomes greater than 1, and the mCP-electron interaction becomes adiabatic. 
This regime would require additional analysis of ionization efficiency, and would also benefit from detectors that are sensitive to energy release, {\em e.g.} phonons, spread between many atoms.)
In Fig. ~\ref{rateinluna} (Right) we illustrate the rate sensitivity and energy thresholds required to probe mCDM with masses below 1 GeV that have accumulated via bottom-up mechanism. Given rapid advance in dark matter detectors, some part of the parameter space can be probed with existing technology. It is also clear that if at some point in the future, the detection thresholds 
for DM-induced recoil can be brought to a sub-eV level (see {\em e.g.} \cite{Knapen:2017xzo}), even $\epsilon$'s as small as $10^{-8}$ can be probed via accelerating mCPs in the underground MV voltage electrostatic accelerators. 

Also of notice is potential sensitivity to mCPs via their generation by cosmic rays which results in an irreducible population on Earth. This irreducible number densities shown in Fig.~\ref{crcap} translate to acceleration rates plotted in Fig.~\ref{ratelunacr}. We find that a maximum obtainable rate can be as high as  $10^3$ Hz, while a lot of new parameter space can be explored with counting rates reaching down to $1$\,Hz and below.

In addition to novel probes discussed in this paper, we have also examined a number of other ways of constraining $\{\epsilon,n_Q\}$ parameter space, summarized in Appendix C. None of them carry as much promise/sensitivity as the three pathways outlined in this work.

\section{Conclusion}
\label{conclusion}

We have shown that the direct probes of milli-charged particles
can be advanced using their accumulation inside the Earth. Owing to a relatively large cross-sections and short free path inside dense media, the number densities of mCPs can indeed be many order of magnitude larger than the cosmological abundances. In this paper we have analyzed the main mechanisms of their accumulation, finding that when evaporation is impossible, strong enhancements of number densities are expected at the locations of the underground laboratories. The enhancement factors can approach $10^{15}$, and therefore even very subdominant fractions of the cosmological mCP DM can be probed. Cosmic ray induced production of mCPs creates far less abundant concentration, which nevertheless could reach up to $10^{-4}$cm$^{3}$.

We pointed out three different methods that could lead to the most precise probes of mCP properties to date. Concentrations upward of $10^{7}$cm$^{-3}$
can be probed via the annihilations of mCP particles to the SM states inside large volume detectors, such as super-Kamiokande. A minimal adjustment of already existing searches will be able to refine sensitivity in 1-to-10 GeV mass range (Fig. 5). mCPs created by cosmic rays cannot be probed this way, as the predicted densities are too small. 

Another method for potentially probing even very small abundances of dark matter is via the formation of bound states of negatively charged mCPs and atomic nuclei. If $\{m_Q,\epsilon\}$ parameters are right ({\em i.e.} just below the solid line on Fig.~\ref{fig4}), the binding to 
heavy elements may be possible, while the binding to light elements is not. In this case, the dark matter experiments that use heavier elements (Xe or I) may be quite sensitive to the energy release accompanying the formation of bound states. In particular, the DM-Ice experiment that utilizes NaI shielded by Antarctic ice is a good example of a device that is extremely sensitive to such a scenario. We argue that $n_Q$ as small as $10^{-7}$cm$^{-3}$ can be probed, that due to accumulation enhancement leads to the sensitivity to tiny $f_Q$. (Detailed analysis of 
$\{m_Q,\epsilon\}$ parameter space is left for future work, as formation of bound states throughout column density may significantly alter the predictions for the density profiles as function of depth.) We also note that for a fixed mCP mass and charge, the amount of binding energy to a given type of an atom is fixed, and therefore the formation of bound states leads to a mono-energetic energy deposition. It is well known that the recently observed Xenon 1T excess events \cite{Aprile:2020tmw} is consistent with a monochromatic signal (see {\em e.g.} \cite{Alonso-Alvarez:2020cdv,An:2020bxd,Baryakhtar:2020rwy,An:2020tcg}). It is then possible to speculate that the excess is maybe coming from $Q^-$-Xe nucleus bound state formation. However, more work needs to be done to understand whether the required densities of the mCPs can be realistically expected for the environment (Gran Sasso lab) where the experiment is operating. 

Finally, perhaps the most direct way of testing the mCP particles is their acceleration in underground accelerators (that see their primary application for measuring astrophysically relevant nuclear reaction cross sections). Even an ``accidental" acceleration of mCPs may result in their kinetic energy going up from thermal to accelerated energies $\sim \epsilon \Delta V$. For MV-type electrostatic accelerators, and for a generous range of $\epsilon$, $10^{-5}-10^{-1}$, the resulting gained kinetic energy can be far above the thresholds of direct detection experiments at 10\,eV (and possibly lower in the near future). Therefore, combining underground accelerations with the specially placed dark matter detectors along the mCP accelerated trajectory can bring significant new sensitivity, and indeed test local concentrations of mCPs down to unprecedented low values. This way, many ``physics targets" can be covered. A dedicated effort in this direction has a potential to explore mCP densities created by cosmic rays, and access the region of parameter space consistent with the explanation of the EDGES anomaly. 

\section{Acknowledgements} We are indebted to 
Drs. A. Berlin and H. Liu for valuable critical comments. 
We are also grateful to Drs. R. Harnik and R. Plestid  for earlier collaboration on the mCP project. M.P. would like to thank Dr. T. Bringmann for earlier discussions of related ideas.  M.P.\ is supported in part by U.S. Department of Energy (Grant No.\ desc0011842).

\appendix
\section{mCP-atom scatterting using Hulthen potential}
For a mCP of mass $M$, and charge $\epsilon$ (with electron charge 1), 
the perturbative expression for an atomic scattering can be written as
\begin{equation}
    \frac{d\sigma}{dq^2}=4\pi Z^2 \frac{\alpha^2 \epsilon^2}{v^2 q^4} F^2(q), 
\end{equation}
where $F(q)$ is atomic form factor normalized to 1 at high-momentum transfer (which corresponds to elastic scattering on un-screened nucleus). 
The same cross section can also be written as,
\begin{equation}
    \frac{d\sigma}{d\Omega}=4Z^2 \frac{\alpha^2 \epsilon^2}{ q^4} F^2(q) 
\end{equation}
We will take the simplest ansatz for $F(q)$ that nevertheless captures the main physics regimes:
\begin{equation}
    F(q)= \frac{a_z^2 q^2}{1+a_z^2q^2}
\end{equation}
where
\begin{equation}
   a_z= \frac{a_0}{4}\left(\frac{9 \pi ^2}{ 2Z}\right)^{\frac{1}{3}}\sim 0.89 \frac{a_0}{Z^{\frac{1}{3}}}\sim \frac{1}{4.2 Z^{\frac{1}{3}} \textrm{keV} }.
\end{equation}

The momentum transfer cross-section is relevant to calculate the overburden required for thermalization as well as the eventual terminal velocity. It is given by
\begin{equation}
    \sigma_T =\frac{\pi}{2(\mu v)^4} \int_0^{2\mu v}\frac{d\sigma}{d\Omega}q^2 dq^2 .
\end{equation}
Setting $a_z \mu v=R$, we get,
\begin{align}
  \sigma_T &=  2\pi\alpha^2 Z^2 \epsilon^2\frac{\log \left(4R^2+1\right)-\frac{4R^2}{4R^2+1}}{\mu^2 v^4} \nonumber \\
  &=2\pi a_z^2{\cal B}^2 \left(\log \left(4R^2+1\right)-\frac{4R^2}{4R^2+1}\right)
\end{align}
Here ${\cal B}=\frac{Z \alpha \epsilon}{a_z \mu v^2}$. It is easy to see that at $R\gg1$ the cross section scales as $v^{-4}$, while in the opposite regime, $R\ll 1$, it does not depend on velocity. 

The perturbative answer is valid only if either $Z \epsilon \alpha a_z  \mu \ll 1$, which for $Z_{\rm rock}\sim 20$ translates to, 
\begin{equation}
    \epsilon \ll 7.5 \times 10^{-5} \frac{\rm GeV}{\mu}
\end{equation}
or if $\frac{Z \epsilon \alpha}{v} \ll 1$, for $Z_{\rm rock}\sim 20$, 
\begin{equation}
    \epsilon \ll 6.3 v
\end{equation}

For larger coupling, expressions are available in the classical limit, $\mu a_z v \gg1$ which is not valid here, $0.6 \sqrt{\frac{\mu}{\rm GeV}} \sim 1$.

We will instead use the expressions derived from Hulthen potential in Ref. \cite{Tulin:2013teo}.
\begin{equation}
\sigma_T=\frac{4\pi}{\mu^2 v^2} \sin^2\delta_0
\end{equation}
Here,
\begin{equation}
    \delta_0 = \textrm{arg} \left(\frac{i \Gamma \left(\frac{i 2\mu a_z v}{\kappa}\right)}{\Gamma(\lambda_+) \Gamma(\lambda_-)}\right)
\end{equation}
\begin{align}
    \lambda_\pm &= 1+\frac{i \mu a_z v}{\kappa}\pm \sqrt{\frac{Z \epsilon \alpha 2\mu}{\kappa m_\phi} - \frac{\mu^2 a_z^2 v^2}{\kappa^2}}\quad \textrm{attractive} \nonumber \\
    &= 1+\frac{i \mu a_z v}{\kappa}\pm i\sqrt{\frac{Z \epsilon \alpha 2\mu}{\kappa m_\phi} + \frac{\mu^2 a_z^2 v^2}{\kappa^2}}\quad \textrm{repulsive}
 \end{align}
 We find that a very good approximation for both attractive and repulsive interactions,
\begin{align}
 \langle \sigma_T \rangle_{\rm th}&\sim \textrm{Min} \left( \frac{16 \pi Z^2 \alpha^2 \epsilon^2}{\mu^2 v_{\rm th}^4} , \frac{4 \pi}{\mu_{\rm rock,Q}^2 v_{\rm th}^2} \right) \nonumber \\
 \langle \sigma_T v\rangle_{\rm th} &\sim  \textrm{Min} \left( \frac{8 \pi Z^2 \alpha^2 \epsilon^2}{\mu^2 v_{\rm th}^4} \times v_{\rm th},   \frac{2.2 \pi}{\mu_{\rm rock,Q}^2 v_{\rm th}^2} \times v_{\rm th} \right) \nonumber \\
\langle \sigma_T v^3\rangle_{\rm th} &\sim   \textrm{Min} \left( \frac{5 \pi Z^2 \alpha^2 \epsilon^2}{\mu^2 v_{\rm th}^4} \times v_{\rm th}^3 ,   \frac{2.2 \pi}{\mu_{\rm rock,Q}^2 v_{\rm th}^2} \times v_{\rm th}^3\right) 
\end{align}
These cross sections form the basis for our code that calculates the effective slow-down, sinking velocity, diffusion coefficients, and ultimately $n_Q(h)$ in the main body of the text.
 
\section{Accelerator geometry}
\label{accgeom}
The dark matter is thermal and hence has velocity 
\begin{equation}
    v_{\rm th}=\sqrt{\frac{2\textrm{T}}{m_Q}}\sim 7\times 10^{-6}\sqrt{\frac{\rm GeV}{m_Q}}
\end{equation}
The differential angular flux coming in per infinitesimal pipe length is
\begin{equation}
    \frac{d\Phi}{dl d\cos{\theta}d\varphi} = 4r  \frac{f_Q \eta \rho_Q}{m_Q}  v_{\rm th}
\end{equation}
Here $\theta\in \{0,\frac{\pi}{2}\}$, is the angle between the incoming particle velocity and the beam axis, while  $\varphi \in \{0,\frac{\pi}{2}\}$ subtended by the velocity on the radial direction.
 
The time spent by the particles inside the pipe is, 
\begin{equation}
 \tau(\theta,\varphi)=   \frac{2r}{v_{\rm th}\sin{\theta}\cos{\varphi}}
\end{equation}

The maximum time the particles can spend (because of acceleration along the beam axis) is,
\begin{equation}
    \tau_{\rm max}=\sqrt\frac{2l} {a}=\sqrt{2 l L} \sqrt{\frac{m_Q} {\epsilon e\Delta V}}
\end{equation}

If a particle spends time $\tau= \textrm{Min}[ \tau_{\rm max},\tau(\theta,\varphi)]$, inside, it is accelerated to,
\begin{equation}
E_Q= \frac{1}{2} m_Q v_f^2=\frac{1}{2} m_Q a^2 \tau^2 
\end{equation}
now, 
\begin{equation}
E_Q[{\rm max}]=\frac{1}{2} m_Q a^2 \tau_{\rm max}^2= \frac{l}{L} \epsilon e \Delta V
\end{equation}

Given a threshold for subsequent detection $E_{\rm thr}$, there is an $l_{\rm min}$, 
\begin{equation}
l_{\rm min}=\frac{E_{\rm thr}}{\epsilon e \Delta V} L
\end{equation}
Furthermore, we require that the particles enter at the correct angle, this is satisfied if,
\begin{equation}
\sin{\theta} \cos{\varphi} < \frac{r \epsilon e \Delta V}{L \sqrt{T E_{\rm thr}}}
\end{equation}
Putting this together we get,
\begin{align}
\Phi[E>E_{\rm thr}]=&\int_{l_{\rm min}}^L dl \int d\cos{\theta} d\varphi \frac{d\Phi}{dl d\cos{\theta}d\varphi} \\ &\times \Theta\left(\frac{r \epsilon e \Delta V}{L \sqrt{T E_{\rm thr}}}-sin{\theta} \cos{\varphi} \right)
\end{align}
This can be simplified further to
\begin{align}
\Phi[E>E_{\rm thr}]=&2\pi r L\left(1-\frac{E_{\rm thr}}{\epsilon e \Delta V} \right)  \frac{f_Q \eta \rho_Q}{m_Q} \\ &\times  v_{\rm th}\textrm{Min}\left[1,\frac{r \epsilon e \Delta V}{L \sqrt{T E_{\rm thr}}}\right].
\end{align}

\section{Other probes of terrestrial milli-charged dark matter}
There are several recent experimental limits which set constraints on a local population of thermalized  DM. We consider them in turn and show that none of them set constraints on terrestrial mCPs.

\textbf{Cryogens} 

Anomalous heating of cryogens was considered in \cite{Neufeld:2018slx,Neufeld:2019xes,Rajendran:2020tmw}. However all of these experiments employ shielding to reduce black-body radiation and hence will cool the mCDM to the shield temperature before they hit the cryogen if the mean-free-path corresponding to the transfer cross-section is smaller than the thickness of the shield. This can be evaluated using 
\begin{equation}
\lambda_\textrm{MFP}=\frac{1}{n_{\rm shield} \sigma_T} \approx 10^{-6} \textrm{cm} \frac{\mu}{\rm GeV}
\end{equation}
Here we approximate $n_{\rm shield} =\frac{10^{22} }{{\rm cm}^3}$ and take the cross-section in Eqn.~\ref{transfer} to be $\sigma_T \approx \frac{4\pi}{\mu^2 v_{\rm th}^2}$. This MFP is much smaller than any realistic shielding. As a result mCDM will not be constrained by cryogens. 

\textbf{LHC beam lifetime} 

In \cite{Neufeld:2018slx}, limits were placed on accumulated DM with contact interactions from LHC beam lifetime, and these arguments can be generalized to other particle storage rings and accelerators.  Rutherford scattering on mCPs can also take particles from the beam. This will come on top of particle loss due to the residual scattering on atomic constituents of the residual gas molecules inside the beam pipe. The scattering on mCP (that can be treated perturbatively in this problem) is at least $\epsilon^2$ times smaller than scattering on electrons and nuclei.  This way, one can estimate the ``best case" sensitivity via comparing the beam losses on atoms vs mCPs
\begin{equation}
n_Q \sim n_{\rm atoms} \epsilon^{-2} \sim 10^{10}\textrm{cm}^{-3}\times (10^{-2}/\epsilon)^2,
\end{equation}
where we took a realistic residual gas density at $10^6$ particles per cm$^3$. 
This may provide some additional sensitivity to mCP if $f_Q$ is large, but it is inferior to other probes discussed in this paper. 

\textbf{Stability of nuclear isomers: $\mathbf{180^\textrm{m}}$ Ta.} 

In \cite{Lehnert:2019tuw,Pospelov:2019vuf} the non-observation of the decay of isomeric Tantalum was used to set limit on terrestrial DM. The form-factor to scatter with  $180^\textrm{m}$ Ta naturally picks mCP masses in the hundreds of GeV and above. For positively charged mCDM, there is Coulomb repulsion with nuclei. The probability of overcoming this is given by the Gamow factor.
\begin{equation}
P_g(T)=e^{-\frac{E_g}{T}}\approx e^{-11 \frac{\epsilon^2}{v_{\rm th}^2}}
\end{equation}
where $E_g=2\mu \left( \pi \alpha Z_{\rm Ta} \epsilon \right)^2$ and $T$ is the ambient temperature. This factor evaluates to tiny values rendering the cross-section too small for \cite{Lehnert:2019tuw} or any future projections of a Tantalum experiment to set relevant limits.  
Negatively charged mCDM of this mass in the heavy mass range can induce $180$m\,$\to180$\,g.s. transition. If the parameter range is such that the formation of bound states with nuclei is possible, de-excitation of 180m isotope will not provide competitive sensitivity to {\em e.g.} nucleus-mCP recombination process discussed in section IV. The main reason for that is relatively small isotopic abundance of 180m that would not be competitive with many kilograms of I or Xe employed by direct detection experiments. If $\epsilon$ is very small (and the mass is in $\sim 100$\,GeV range to ensure efficient de-excitation), then surface direct detection experiments will be more sensitive to $Q^-$ than Tantalum 180m. 

\textbf{Anomalous heat transport} 
The heat conductivity of Earth can be modified by the presence of an Earth-bound exotic species \cite{Neufeld:2018slx}. However, the heat conductivity is proportional to the mean-free-path which is extremely small in rock as explained above. As a result, there are no useful limits that we can derive from this effect.   
\bibliographystyle{unsrt}
\bibliography{biblio}
\end{document}